\documentclass[twocolumn,superscriptaddress,amsmath,amssymb,aps,prb]{revtex4-1}

\usepackage{graphicx}
\usepackage{dcolumn}
\usepackage{bm}
\usepackage[colorlinks,citecolor=blue,linkcolor=blue,urlcolor=blue,plainpages=false,pdfpagelabels]{hyperref}

\newcommand{\ket}[1]{\left| #1 \right\rangle}

\newcommand{\bra}[1]{\left\langle #1 \right|}

\newcommand{\hc}{{\text{h.c.}}}

\begin{document}

\preprint{APS/123-QED}

\title{Transport properties of Majorana bound states networks\\ in the Coulomb blockade regime}

\author{Johan Ekstr\"om}
\email{johan.ekstrom@uni.lu}
\affiliation{Physics and Materials Science Research Unit, University of Luxembourg, L-1511 Luxembourg}

\author{Patrik Recher}%
%\email{p.recher@tu-braunschweig.de}
\affiliation{Institute of Mathematical Physics, Technical University of Braunschweig, D-38106 Braunschweig, Germany}
\affiliation{Laboratory for Emerging Nanometrology Braunschweig, D-38106 Braunschweig, Germany}

\author{Thomas L. Schmidt}
%\email{thomas.schmidt@uni.lu}
\affiliation{Physics and Materials Science Research Unit, University of Luxembourg, L-1511 Luxembourg}

\begin{abstract}
Topologically protected qubits based on nanostructures hosting Majorana bound states (MBSs) hold great promise for fault-tolerant quantum computing. We study the transport properties of nanowire networks hosting MBSs with a focus on the effects of the charging energy and the overlap between neighboring MBSs in short mesoscopic samples. In particular, we investigate structures hosting four MBSs such as T-junctions and Majorana boxes. Using a master equation in the Markovian approximation, we discuss the leading transport processes mediated by the MBSs. Single-electron tunneling and processes involving creation and annihilation of Cooper pairs dominate in the sequential tunneling limit. In the cotunneling regime the charge in the MBSs network is fixed and transport is governed by transitions via virtual intermediate states. Our results show that four-terminal measurements in the T-junction and Majorana box geometries can be useful tools for the characterization of the properties of MBSs with finite overlaps and charging energy.
\end{abstract}

\maketitle

\section{Introduction} \label{sec:level1}

The transport properties of Majorana bound states (MBSs) in condensed-matter systems have been a flourishing research area in recent years \cite{leijnse_introduction_2012,beenakker_search_2013,read_paired_2000}. This activity is partially motivated by several recent proposals for topologically protected qubits based on MBSs as building blocks, which according to theory could allow for more fault tolerant quantum computation architectures \cite{kitaev_unpaired_2001,alicea_non-abelian_2011,sarma_majorana_2015,terhal_majorana_2012}. One of the most important of such host systems for MBSs is a nanowire with strong spin-orbit coupling, in proximity to a superconductor and subjected to a magnetic field in the direction of the wire \cite{oreg10,leijnse_introduction_2012}. Recent experiments have indeed shown transport signatures consistent with theoretical expectations for transport through MBSs \cite{mourik_signatures_2012, deng_majorana_2016, albrecht_exponential_2016, zhang18}.

As MBSs owe their topological protection to the conservation of fermion parity, a single MBS-based qubit typically involves at least four MBSs. Potential structures for such nanowire-based qubits are the T-junction \cite{weithofer_electron_2014,alicea_non-abelian_2011,clarke_majorana_2011,sau_controlling_2011} and the Majorana box \cite{beri_topological_2012,beri_majorana-klein_2013,altland_multiterminal_2013,altland_multichannel_2014,gau_quantum_2018,zazunov_6ensuremathpi_2017}. Both systems consist of nanowires placed on top of a mesoscopic superconductor and can be controllably connected to metallic leads. The T-junction and the Majorana box have been suggested for different purposes. The T-junction is one of the simplest geometries where MBSs can be braided and their non-Abelian exchange statistics can be detected \cite{alicea_non-abelian_2011,clarke_majorana_2011,sau_controlling_2011,van_Heck_2012}. The Majorana box, on the other hand, has been proposed as part of a two-dimensional lattice for implementing Majorana surface codes \cite{schmidt_2008,terhal_2012,plugge_majorana_2016}.

The manipulation of such qubits also requires read-out processes to be able to keep track of the state of the qubit \cite{aasen_milstones_2016,plugge_majorana_2016}. It is therefore crucial to understand the transport properties of these structures. Single wires hosting a pair of MBSs have already been studied in detail \cite{bolech_observing_2007,fu_electron_2010,zazunov_coulomb_2011,hutzen_majorana_2012,van_heck_conductance_2016}. It was recognized that the mesoscopic size of the superconductor, and the associated charging energy of a ``floating'', as opposed to a grounded, superconductor can lead to interesting effects such as quantum teleportation \cite{fu_electron_2010}. Moreover, it was shown that the unavoidable hybridization of MBSs in finite-size systems also yields distinct signatures \cite{dassarma12}.

In contrast to single wires, both the Majorana box and the T-junction hold four MBSs and thus allow for new types of transport processes. For instance, it was shown that a T-junction placed on a grounded superconductor can exhibit double crossed Andreev reflection \cite{weithofer_electron_2014}. This transport process is due to concurring resonant Andreev reflections on the outer leads and a non-resonant process on the central lead.

The Majorana box has been investigated in greater detail than the T-junction. When there is no overlap between the MBSs the system can be mapped to a degenerate spin-1/2 system, which leads to a ``topological Kondo effect'' at temperatures below a Kondo temperature \cite{beri_topological_2012, beri_majorana-klein_2013, altland_multichannel_2014,two-channel_Landau_2017}. Gau \textit{et al.} have recently presented a very detailed analysis on the transport properties in systems with multiple Majorana boxes \cite{gau_quantum_2018}.

In this paper we consider the T-junction and the Majorana box and set out to improve the understanding of their transport properties when both the charging energy of the mesoscopic floating superconductor and their mutual overlaps are important. We begin by writing down a general theory, applicable to nanowire networks, based on the master equation and rate equations in the weak tunneling limit. We then apply it to the T-junction and the Majorana box. For resonant transport, i.e., in the sequential tunneling limit, we find transport features which can be interpreted as a generalization of the quantum teleportation process proposed for single Majorana wires \cite{fu_electron_2010}. Other features can be regarded as a form of nonlocal transport between two Majorana wires, via the Cooper pairs of the superconductor. Away from the resonances, we study the cotunneling regime, where fluctuations of the particle number are suppressed due to the charging energy and transport occurs solely via virtual states.

This paper is organized as follows: first, we present the general model for a nanowire network hosting MBSs, see Sec.~\ref{Sec: Model}. Afterwards, we will discuss the specific details of the T-junction and the Majorana box. In Sec.~\ref{Sec: SeqT}, we present the theory and the results of the transport in the sequential tunneling limit. In Sec.~\ref{Sec: CoT}, the transport in the cotunneling limit is explored. We conclude in Sec.~\ref{Sec: Conclusions}. Throughout this paper we let $\hbar = e = k_B = 1$.

\section{Model} \label{Sec: Model}

We begin this section by describing a general system consisting of a network of nanowires hosting MBSs, placed on top of a floating mesoscopic superconductor. The system is coupled to a gate electrode which allows for control over the electrostatic energy of the superconducting island. This charging energy is given by $E_{C}(2N_{C} + n + n_{g})^2$, where $E_{C}$ is related to the electrostatic capacity of the island, $N_{C}$ is the number of Cooper pairs in the superconductor, and $n$ is the number of electrons in the MBSs of the nanowire-network.

\begin{figure}
\centering
\includegraphics[trim=5cm 15cm 6cm 4cm, clip,scale=0.4]{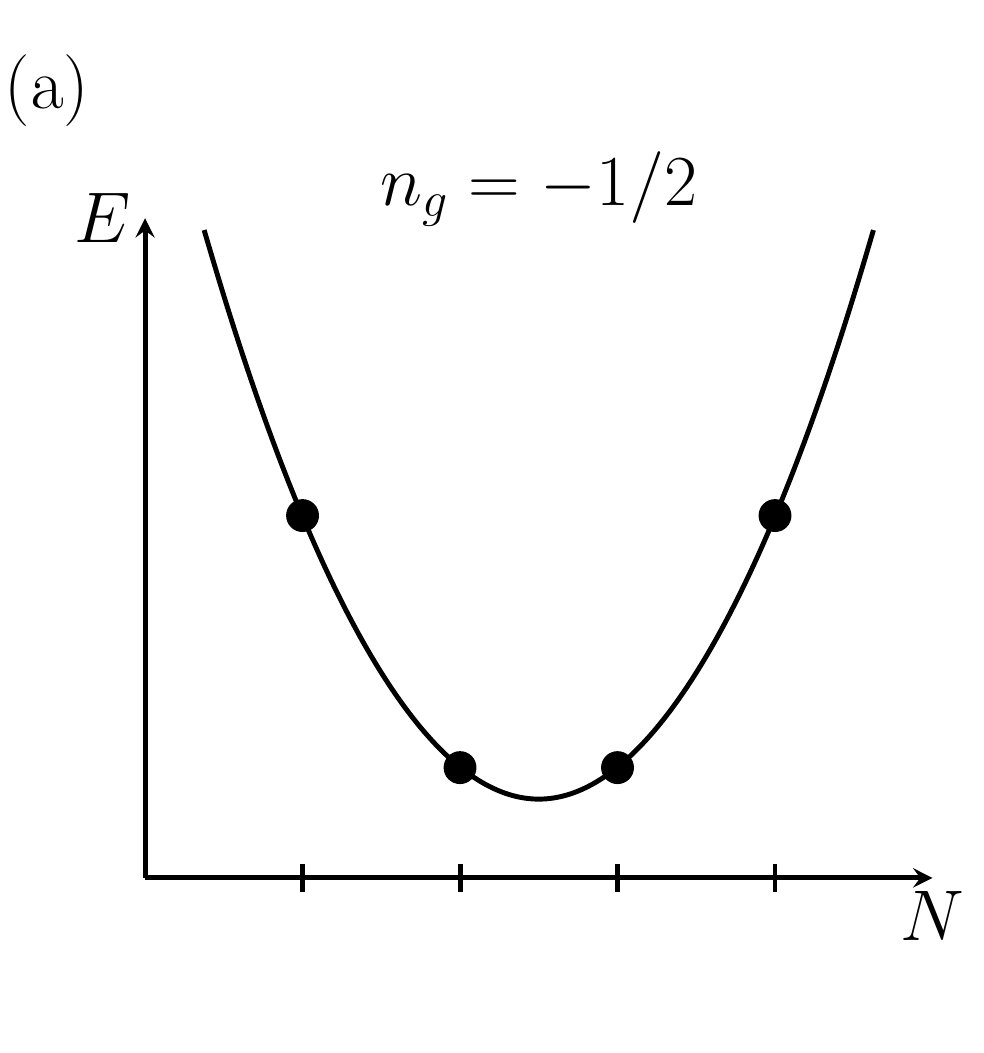}
\includegraphics[trim=5cm 15cm 6cm 4cm, clip,scale=0.4]{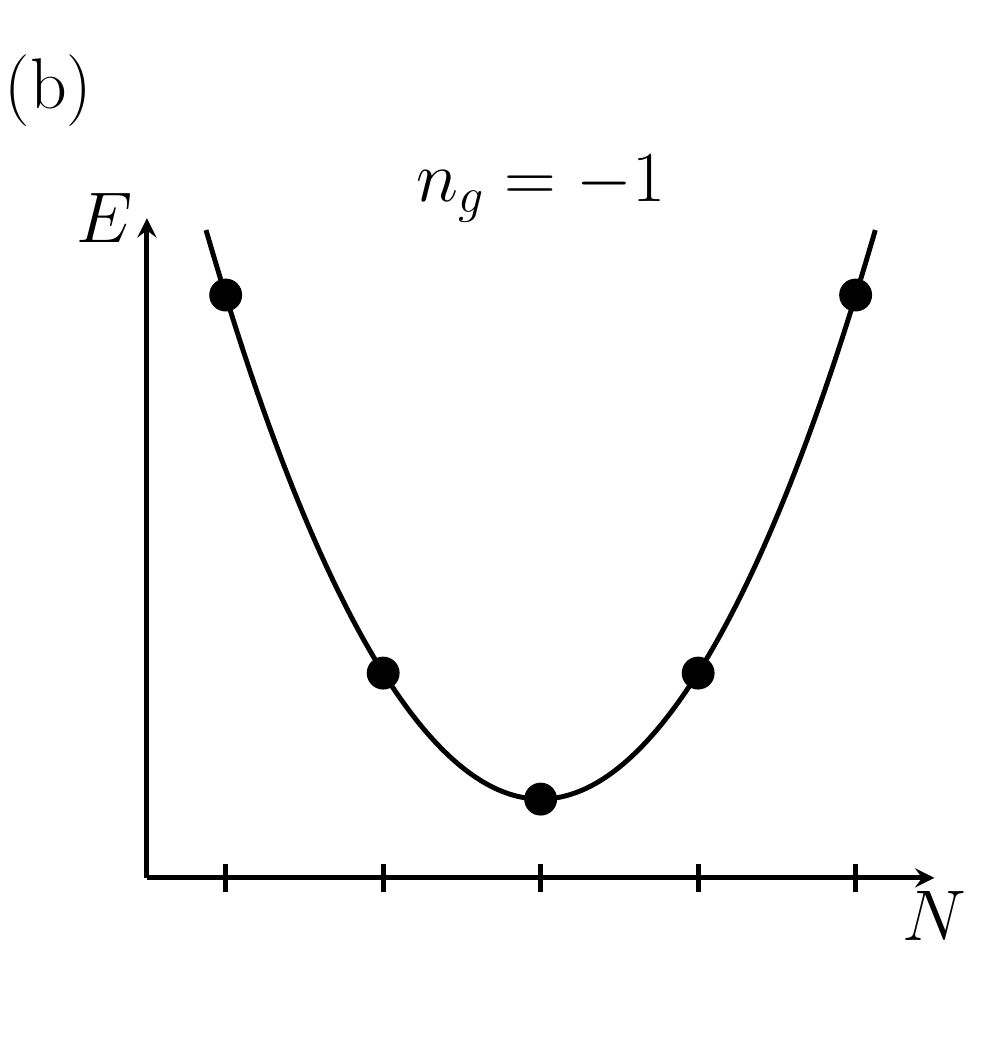}
\caption{Charging energy of the superconducting island as a function of the total number of electrons $N = 2N_{C} + n$. By tuning the gate voltage the system can reach the two different regimes depicted in the figure. (a) For $n_{g} = -1/2 + k$ with $k\in \mathbb{Z}$ there are two degenerate charge states at the bottom of the parabola. In this regime sequential tunneling is the dominant transport mechanism. (b) In contrast, when $n_{g} = -1 + k$ a single charge state is at the bottom of the parabola. The leading transport process is then cotunneling due to transitions via virtual states.}
\label{fig: parabolas}
\end{figure}

The parameter $n_{g}$ can be adjusted via the gate voltage, and can be used to tune the system into two distinct transport regimes, as depicted in Fig.~\ref{fig: parabolas}. For values $n_{g} = -1/2 + k$ with $k\in \mathbb{Z}$, degenerate charge states are obtained and sequential tunneling via the degenerate ground state is the dominant process. A different regime is reached for $n_{g} = -1 + k$. A single charge state sits at the bottom of the parabola depicted in Fig.~\ref{fig: parabolas}(b), so the ground state is non-degenerate. In this regime cotunneling is the leading process for transport, and involves transport via virtual states.

We will consider the parameter regime $\Delta \gg E_{C}, |\mu_l|$, where $\Delta$ is the superconducting gap and $\mu_l$ is the chemical potential of lead $l$ measured from the center of the superconducting gap. In this regime quasiparticle excitations in the superconductor can be neglected and only the MBSs need to be taken into account to describe the low-energy physics of the system. The Hamiltonian describing the $2M$ MBSs in the network is
\begin{equation}
H_{\rm MBS} = -i\sum_{k=1}^{2M} \sum_{l = k+1}^{2M} \epsilon_{kl}\gamma_{k}\gamma_{l},
\label{Eq: HMBS}
\end{equation}
where $\epsilon_{kl}$ is the energy splitting due to the overlap of the wave functions of the MBSs $\gamma_k$ and $\gamma_l$. They fulfill the anticommutation relations $\lbrace \gamma_{l},\gamma_{k}\rbrace = 2\delta_{lk}$. The network is connected to $2M$ metallic leads which are described by effective one-dimensional Hamiltonians,
\begin{align}
H_{\rm leads} = -iv_{F}\sum_{l=1}^{2M}\int dx\psi_{l}^{\dagger}(x)\partial_{x}\psi_{l}(x),
\end{align}
where $\psi^{(\dagger)}_{l}(x)$ is the annihilation (creation) operator in lead $l$ and $v_{F}$ is the Fermi velocity. As we assume electron tunneling to be local and focus on low energies, the leads can be approximated as one-dimensional systems with a constant density of states. Moreover, the leads can be considered as spinless electron reservoirs because only one spin orientation will couple to the MBS \cite{bolech_observing_2007}. Tunneling of electrons between the leads and the superconducting island is described by \cite{bolech_observing_2007}
\begin{equation}
H_{\rm tun} = \sum_{l=1}^{2M} \ t_{l}\psi_{l}^{\dagger}(x=0) \gamma_{l} + \hc ,
\label{Eq: HT1}
\end{equation}
where $t_{l}$ is the tunneling amplitude between lead $l$ and the MBS $\gamma_l$ which it is connected to. Next, we transform from the Majorana basis to a Dirac basis. For every pair of MBSs in the system we construct a Dirac fermion $c_j$ such that (for $j = 1, \ldots, M$),
\begin{align}
\gamma_{2j-1} &= c_{j} + c_{j}^{\dagger}, \nonumber \\
\gamma_{2j} &= i(c_{j}^{\dagger} - c_{j}).
\label{Eq: Maj12}
\end{align}
We insert Eq.~(\ref{Eq: Maj12}) into Eq.~(\ref{Eq: HMBS}) and diagonalize the resulting Hamiltonian using a Bogoliubov transformation (see App.~\ref{Apn: BTT}). Hence, one obtains the representation
\begin{equation}
H_{\rm MBS} = \sum_{j=1}^M \xi_{j}d_{j}^{\dagger}d_{j},
\label{Dirac}
\end{equation}
where $d_j$ are Dirac fermionic operators and $\xi_{j}$ are the corresponding eigenenergies. The MBSs can be expressed in terms of these as
\begin{align}
\gamma_{l} = \sum_{j=1}^M \left( \alpha_{lj}d_{j} + \alpha_{lj}^{*}d_{j}^{\dagger} \right),
\label{Eq: M=D}
\end{align}
where $\alpha_{lj}$ are the elements of a $2M \times M$ matrix obtained from the Bogoliubov transformation. By inserting Eq.~(\ref{Eq: M=D}) into Eq.~(\ref{Eq: HT1}) we obtain the tunneling Hamiltonian in terms of Dirac operators,
\begin{equation}
H_{\rm tun} =  \sum_{l=1}^{2M} \sum_{j = 1}^M t_{l}\psi_{l}^{\dagger}(x=0) \left( \alpha_{lj} d_{j} + \alpha_{lj}^{*} d_{j}^{\dagger}\right) + \hc
\label{Eq: HTD}
\end{equation}

When the system Hamiltonian is expressed in terms of Dirac operators the electrostatic energy can easily be taken into account. It is described by
\begin{equation}
H_{\rm charging} = E_{C}\left(\hat{n} + n_{g} + 2\hat{N}_{C}\right)^2.
\end{equation}
Here, $\hat{n} = \sum_{j} d^{\dagger}_{j}d_{j}$ is the number operator for the Dirac fermions in the MBSs and $\smash{\hat{N}_{C}}$ is the number operator for the Cooper pairs. Note that since the Bogoliubov transformation is unitary, $\hat{n}$ is basis-independent.

To study electron transport, it is convenient to rewrite the tunneling Hamiltonian in a charge conserving way. This can be achieved by inserting the operators $e^{\pm i\phi}$, where $\phi$ denotes the operator for the superconducting phase. Since $\hat{N}_{C}$ is conjugate to the superconducting phase $\phi$, $[ \phi, \hat{N}_{C} ] = i$, its exponentials $e^{\pm i\phi}$ increase or decrease, respectively, the number of Cooper pairs on the superconducting island by one \cite{fu_electron_2010}. As the operators $\psi^{\dagger}_l$ and $d^{\dagger}_j$ both create one unit of charge, we make the tunneling Hamiltonian charge-conserving by using
\begin{equation}
H'_{\rm tun} =  \sum_{l=1}^{2M} \sum_{j = 1}^M  t_{l}\psi_{l}^{\dagger}(x=0) \left( \alpha_{lj} d_{j} + \alpha_{lj}^{*} d_{j}^{\dagger}e^{-i\phi}\right) + \hc
\label{Eq: HTD2}
\end{equation}
$H'_{\rm tun}$ contains two terms and their hermitian conjugates. The first one describes normal single-electron tunneling where an electron is annihilated on the island while another electron is created in the lead. The second term describes an Andreev process where two electrons are created, one in the lead and one in the wire, while a Cooper pair is annihilated in the superconductor.

The total Hamiltonian is given by
\begin{align}
H = H_{\rm leads} + H_{\rm MBS} + H_{\rm charging} + H'_{\rm tun}.
\end{align}
In the following, we will consider $H'_{\rm tun}$ as a perturbation. The Hilbert space of the island Hamiltonian $H_{\rm MBS} + H_{\rm charging}$ is spanned by the states $\ket{\mathbf{n};N_{C}}$, where $\mathbf{n} = ( n_1 , \ldots, n_M)$ and $n_{j} \in \{0,1\}$, denotes whether the Dirac state $d_j$ of the island is empty or occupied. The leads are considered to be Fermi seas at chemical potential $\mu_l$. We have now presented the general setup and will, in the next two subsections, give the specific descriptions of the Majorana box and the T-junction.

\subsubsection{Majorana box}
A schematic picture of the Majorana box is shown in Fig.~\ref{fig: setup}(a). Two wires are placed on the same superconducting island and only the MBSs on the same wire overlap with each other. The Hamiltonian describing the MBSs in the Majorana box is therefore
\begin{align}
H_{\text{Box}} = -i\epsilon_{12}\gamma_{1}\gamma_{2} - i \epsilon_{34}\gamma_{3}\gamma_{4}.
\end{align}
The MBSs on different wires interact with each other via the charging energy. By using Eq.~(\ref{Eq: Maj12}), the Majorana box Hamiltonian already becomes diagonal in terms of fermionic operators. Therefore, the Bogoliubov transformation is trivial ($c_{i} \equiv d_{i}$) and one finds,
\begin{equation}
H_{\text{Box}} = \xi_{1}d^{\dagger}_{1}d_{1} + \xi_{2} d_{2}^{\dagger}d_{2},
\end{equation}
with $\xi_{1} = \epsilon_{12}$ and $\xi_{2} = \epsilon_{34}$. The coupling coefficients $\alpha_{lj}$ for the Majorana box read as follows
\begin{align}\label{Tab: Mbox}
\begin{array}[t]{|c|c|c|}
\hline
\alpha_{lj} & j=1 & j=2 \\
\hline
l=1 & 1 & 0 \\
l=2 & i & 0 \\
l=3 & 0 & 1 \\
l=4 & 0 & i \\
\hline
\end{array}
\end{align}
As expected one finds that leads $1$ and $2$ only couple to the $d_{1}$ mode, whereas leads $3$ and $4$ only couple to the $d_{2}$ mode.

\subsubsection{T-junction}
\begin{figure}
\centering
\includegraphics[trim=9.4cm 7cm 1cm 4cm, clip,scale=0.4]{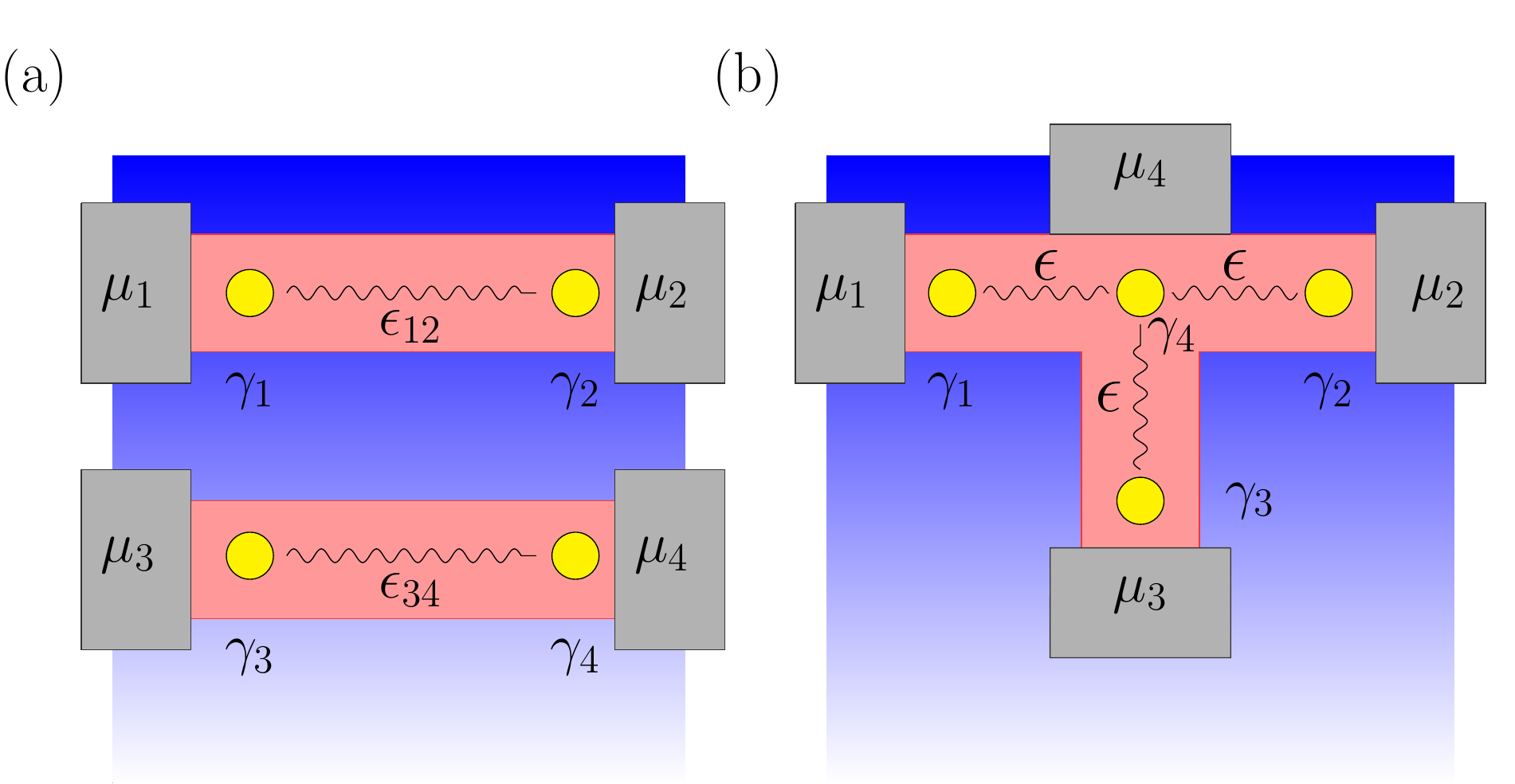}
\caption{Schematic picture of (a) the Majorana box and (b) the T-junction. The blue area represents the superconductor on which the nanowire network (in red) is placed. Yellow circles represent the MBSs $\gamma_{l}$ and the wiggly lines the overlap between them. Due to the exponential decay of the MBS wave functions only overlap between nearest neighbors is considered. Gray regions represent the metallic leads at chemical potentials $\mu_l$. Each Majorana is tunnel-coupled to the corresponding lead.}
\label{fig: setup}
\end{figure}

The T-junction consists of two crossed nanowires that form a T-shaped structure, see Fig.~\ref{fig: setup}(b) \cite{plissard_formation_2013}. The MBS wave functions decay exponentially into the nanowires, so assuming the nanowires to be long enough, only the overlap between nearest neighbor MBSs needs to be considered \cite{dassarma12}. This results in the following Hamiltonian,
\begin{align}
H_{\text{T-junction}} = -i\epsilon\gamma_{1}\gamma_{4} -i \epsilon\gamma_{2}\gamma_{4} -i \epsilon\gamma_{3}\gamma_{4}. \label{eq:T}
\end{align}
Diagonalizing this Hamiltonian leads to two linear combinations of MBSs at a finite energy $\xi_{T} = 2 \sqrt{3} \epsilon$ and two linear combinations that remain at zero energy. Using a Bogoliubov transformation (see App.~\ref{Apn: BTT} for details), the diagonalized Hamiltonian becomes
\begin{equation}
H_{\text{T-junction}} = 0 d^{\dagger}_{1}d_{1} + \xi_{T}d^{\dagger}_{2}d_{2},
\label{Eq: TJDirac}
\end{equation}
with the two mutually anticommuting fermionic operators,
\begin{align}
d_{1} &= \frac{1}{4\sqrt{3}}\left[(1+\sqrt{3}) + (1-\sqrt{3})i\right]\gamma_{1} \notag \\
&+ \frac{1}{4\sqrt{3}}\left[(1-\sqrt{3}) + (1+\sqrt{3})i\right]\gamma_{2} \notag \\
&- \frac{1}{2\sqrt{3}}(1+i)\gamma_{3}, \label{Eq: d1T} \\
d_{2} &= \frac{1}{2\sqrt{3}}\left[\gamma_{1} + \gamma_{2} + \gamma_{3} - \sqrt{3}i\gamma_{4}\right]. \label{Eq: d2T}
\end{align}
By inverting these expressions we obtain the coefficients $\alpha_{lj}$ describing the coupling between the Dirac states and the leads used in Eq.~(\ref{Eq: HTD}). Explicitly, they are given by
\begin{align}\label{Tab: Tjunction}
\begin{array}{|c|c|c|}
\hline
\alpha_{lj} & j=1 & j=2 \\
\hline
l=1 & \sqrt{\frac{2}{3}}e^{i\pi/12} & \frac{1}{\sqrt{3}} \\
l=2 & \sqrt{\frac{2}{3}}e^{-7i\pi/12} & \frac{1}{\sqrt{3}} \\
l=3 & \sqrt{\frac{2}{3}}e^{3i\pi/4} & \frac{1}{\sqrt{3}} \\
l=4 & 0 & i \\
\hline
\end{array}
\end{align}
The Hamiltonian of the T-junction thus has a Dirac state $d_2$ at energy $\xi_{T}$ and a two-fold degenerate state $d_1$ at zero energy. The $d_{2}$ mode is a linear combination of all four MBSs of the system, whereas the $d_{1}$ mode is a nonlocal state involving only the three outer MBSs. Table~(\ref{Tab: Tjunction}) also reveals that the central lead does not couple to the $d_{1}$ mode as $\alpha_{41} = 0$. Note that the fact that $\alpha_{41} = 0$ is not an artifact of the assumed symmetry between the overlaps of the outer MBSs with the central one. Indeed, a nonzero $\alpha_{41}$ would only arise if next-nearest neighbor overlaps were considered in Eq.~(\ref{eq:T}).

\section{Transport} \label{Sec: Transport}

In this section, we present the theory used to calculate the transport processes in the sequential tunneling regime and the cotunneling regime. The starting point is a master equation from which we then obtain the average current and the differential conductance. We consider the weak-coupling limit, where the tunneling rate $\Gamma_l$ from lead $l$ to the island is small compared to either the temperature or the chemical potentials of the leads. To observe Coulomb blockade we further assume that $T \ll E_{C}$ where $E_{C}$ is the charging energy \cite{nazarov_blanter_2009}.

\subsection{Sequential tunneling} \label{Sec: SeqT}

To calculate the current we first introduce counting operators in the tunneling Hamiltonian. This is a necessary step since the Born-Markov approximation involves tracing out the lead degrees of freedom. For this purpose, we define for each lead a number operator $\hat{N}_{l}$, which is considered to be part of the reduced system, and which counts the number of particles that have tunneled to lead $l$. It is canonically conjugate to a lowering operator $Y_{l}$ (raising operator $Y^{\dagger}_{l}$) that creates (annihilates) an electron in lead $l$,
\begin{align}
\big[\hat{N}_{l},Y_{l}^{}\big] &= -Y_{l}, \label{CF1} \\
\left[\hat{N}_{l},Y_{l}^{\dagger}\right] &= Y_{l}^{\dagger}.
\label{CF2}
\end{align}
Including the counting operators, the tunneling Hamiltonian can now be written as
\begin{equation}
H''_{\rm tun} =  \sum_{l=1}^{2M} \sum_{j = 1}^M  t_{l}\psi_{l}^{\dagger}(x=0)Y^{\dagger}_l \left( \alpha_{lj} d_{j} + \alpha_{lj}^{*} d_{j}^{\dagger}e^{-i\phi}\right) + \hc
\label{Eq: HTD3}
\end{equation}
To calculate the current and conductance in the sequential tunneling regime we can use the following master equation in the Born-Markov approximation \cite{breuer_theory_2002},
\begin{align}
& \frac{d}{dt}\rho_{S}(t) \label{Eq: Master1}\\
& =  \int_{0}^{\infty} ds\hspace{1pt} \text{Tr}_{B}\lbrace \left[H''_{\rm tun}(t),\left[\rho_{S}(t)\otimes\rho_{B},H''_{\rm tun}(t-s)\right]\right]\rbrace,
\notag
\end{align}
which describes the time evolution in the interaction picture of the reduced density matrix $\rho_S(t)$ of the system consisting of the superconducting island and the MBSs. The lead degrees of freedom, in contrast, form a fermionic bath and have been traced over.

We are interested in the current in the stationary limit. Because fast oscillating terms average out in the stationary limit, we make a secular approximation and neglect them. This leads to a set of equations for the populations $P(\mathbf{n};N_{C}) = \bra{\mathbf{n};N_{C}} \text{Tr}_{\hat{N}_l} \rho_S \ket{\mathbf{n};N_{C}}$. We look for stationary solutions, i.e., $\frac{d}{dt}P(\mathbf{n};N_{C}) = 0$, with the condition that the sum of occupation probabilities is one. The resulting coupled equations for the corresponding occupation probabilities can thus easily be solved numerically.

From this the average current in lead $l$ can then be calculated from
\begin{align}
\langle I_{l} \rangle &= \text{Tr}\left\lbrace\hat{N}_{l}\frac{d}{dt}\rho_{S}\right\rbrace \notag \\
&= \sum_{\mathbf{n},N_{C}}\left( \Gamma_{l}^{1,+} - \Gamma_{l}^{1,-} \right) P(\mathbf{n},N_{C}),
\label{Eq: CurrentSQ}
\end{align}
where $\Gamma_{l}^{1,+(-)}$ denote transition probabilities resulting from processes that increase (decrease) the number of electrons in lead $l$ by one. Explicitly, they are given by
\begin{widetext}
\begin{align}
\Gamma^{1,+}_{l} &= \Gamma_{l}|\alpha_{li}|^{2}\Big\lbrace 2 - n_{F}\Big[ - \xi(\mathbf{n}) - E_{C}\Big(1 - 2n_{g} - 4N_{C}\Big) - \mu_{l}\Big] - n_{F}\Big[ \xi(\mathbf{n}) + E_{C}\Big(1 - 2n_{g} - 4N_{C}\Big) - \mu_{l} \Big] \Big\rbrace,
\\
\Gamma^{1,-}_{l} &= \Gamma_{l}|\alpha_{li}|^{2}\Big\lbrace n_{F}\Big[ - \xi(\mathbf{n}) - E_{C}\Big(1 - 2n_{g} - 4(N_{C}+1)\Big) - \mu_{l}\Big] + n_{F}\Big[ \xi(\mathbf{n}) + E_{C}\Big(1 - 2n_{g} - 4N_{C}\Big)  - \mu_{l}\Big] \Big\rbrace.
\end{align}
\end{widetext}
Here, $\Gamma_{l} = |t_{l}|^{2}/v_{F}$, $n_F(\omega) = 1/(e^{\omega/T} + 1)$ is the Fermi function, and
\begin{align}
    \xi(\mathbf{n}) = \sum_j \xi_j n_j
\end{align}
is the energy of the state $\ket{\mathbf{n}}$. The local or nonlocal differential conductances can be calculated from $d\langle I_l \rangle /d\mu_k$. Due to the finite charging energy, only a small number of states is accessible in the bias window and thus contributes to transport. In particular, whether or not tunneling is possible in the sequential tunneling regime, depends on degeneracies of the energy of the reduced system,
\begin{align}
    E(\mathbf{n}, N_C) = \xi(\mathbf{n}) + E_C \bigg( \sum_j n_j + n_g + 2 N_C \bigg)^2.
\end{align}
Without loss of generality, we will use $N_C = 0$ and $-1 < n_g < 0$ in the ensuing discussion. Moreover, to illustrate the transport regimes, we will assume $\xi_j \geq 0$.

\subsubsection{Majorana box}

\begin{figure}[t]
\includegraphics[scale=0.26]{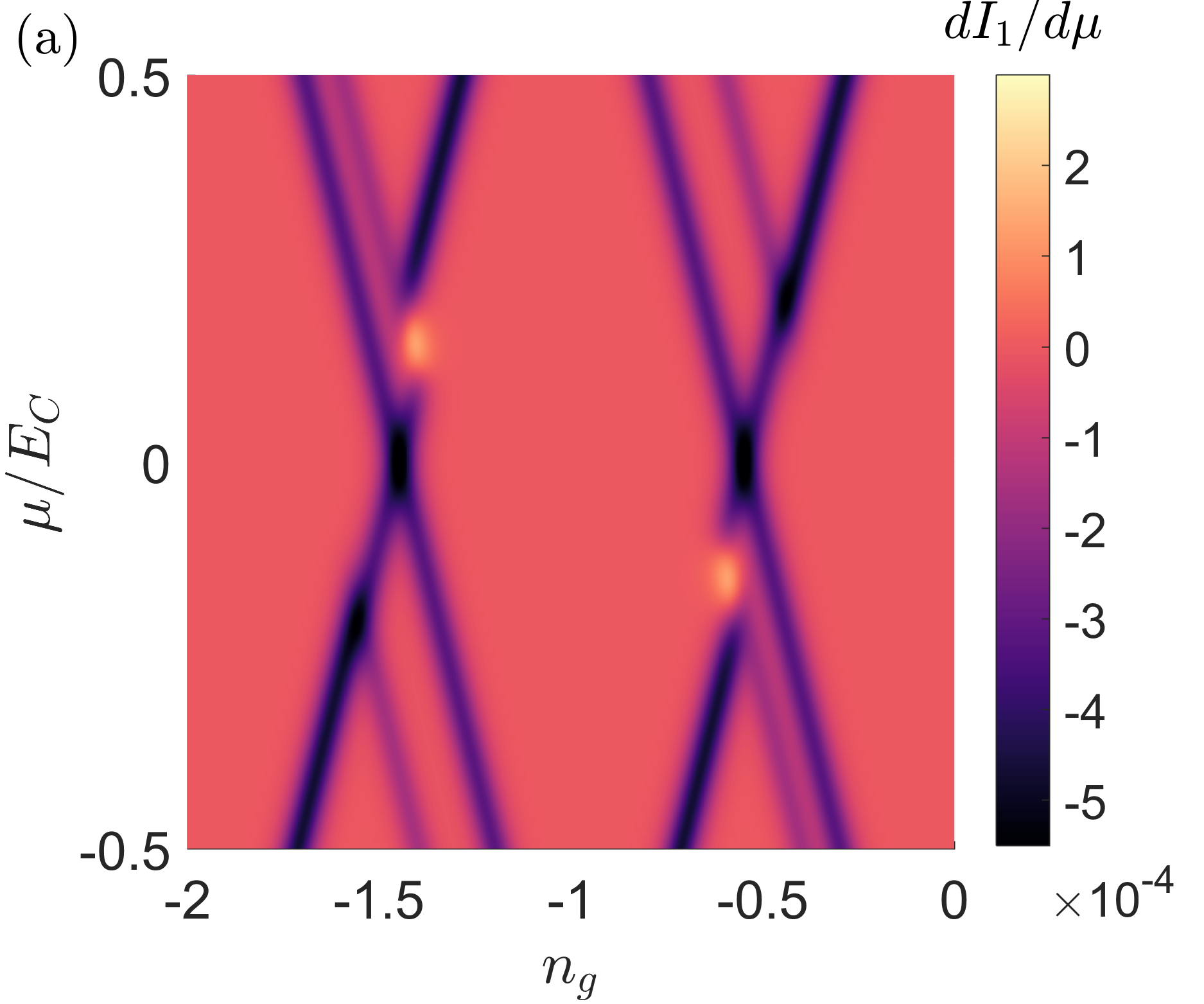}
\includegraphics[scale=0.26]{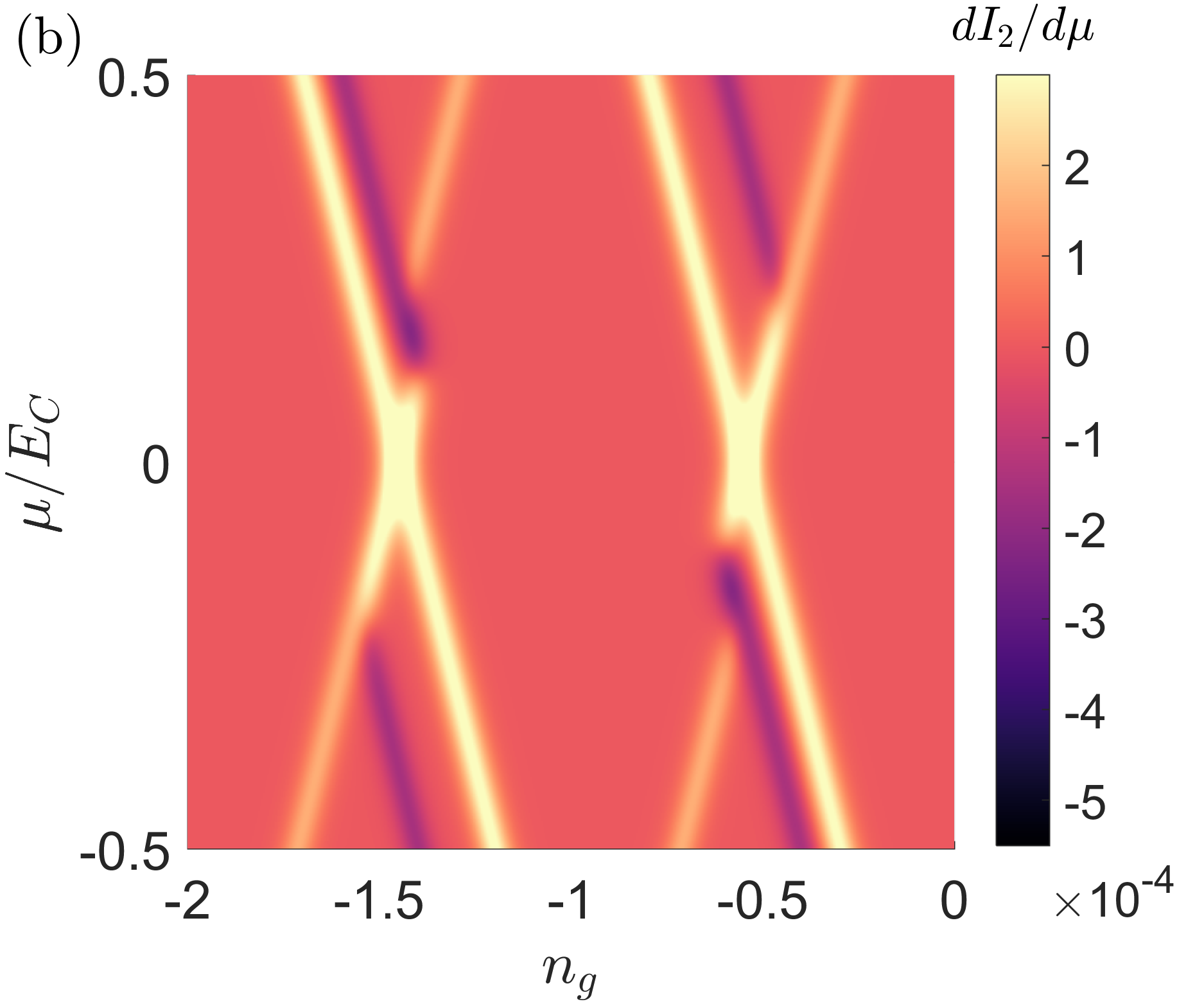}\\
\includegraphics[scale=0.26]{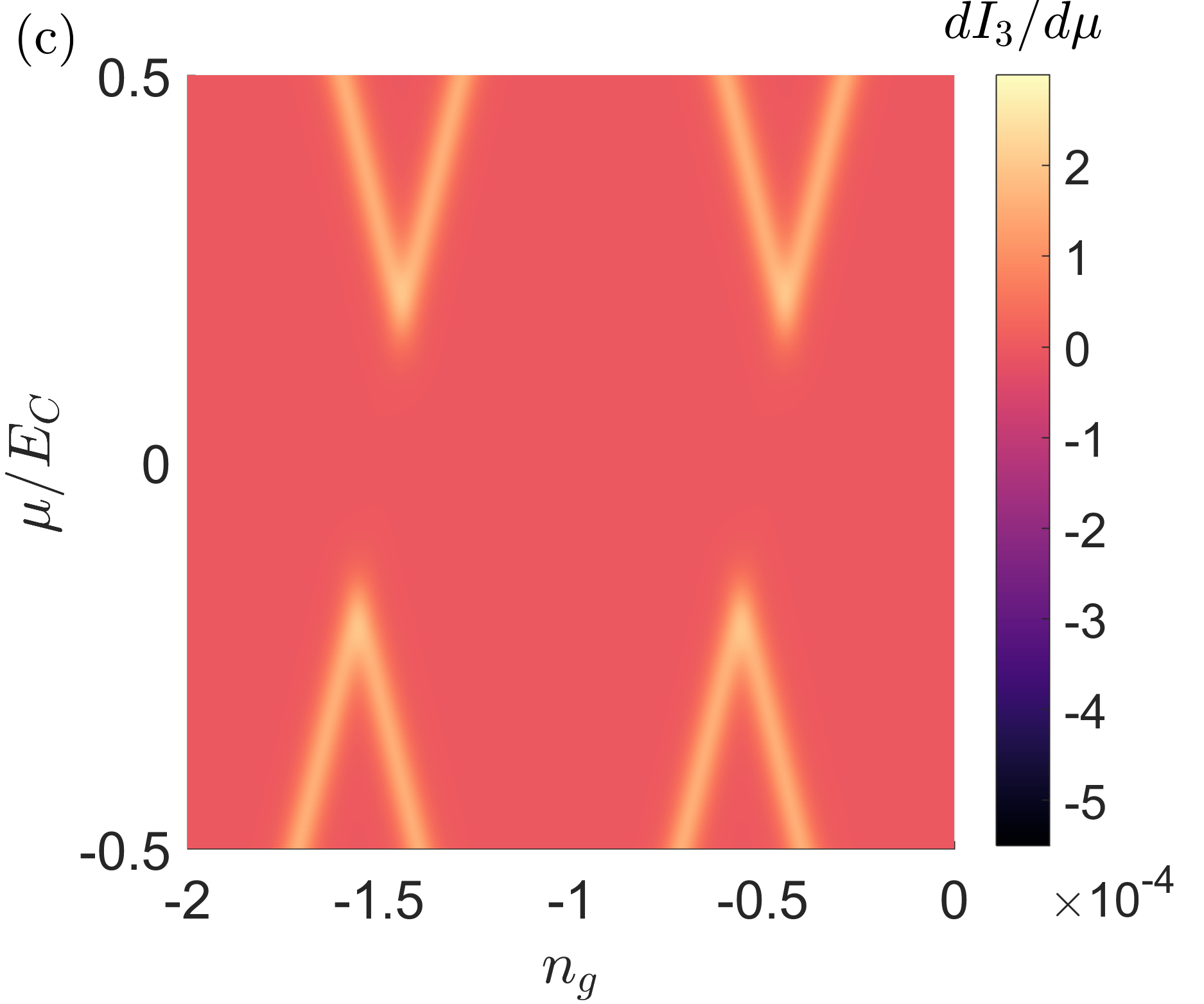}
\includegraphics[scale=0.26]{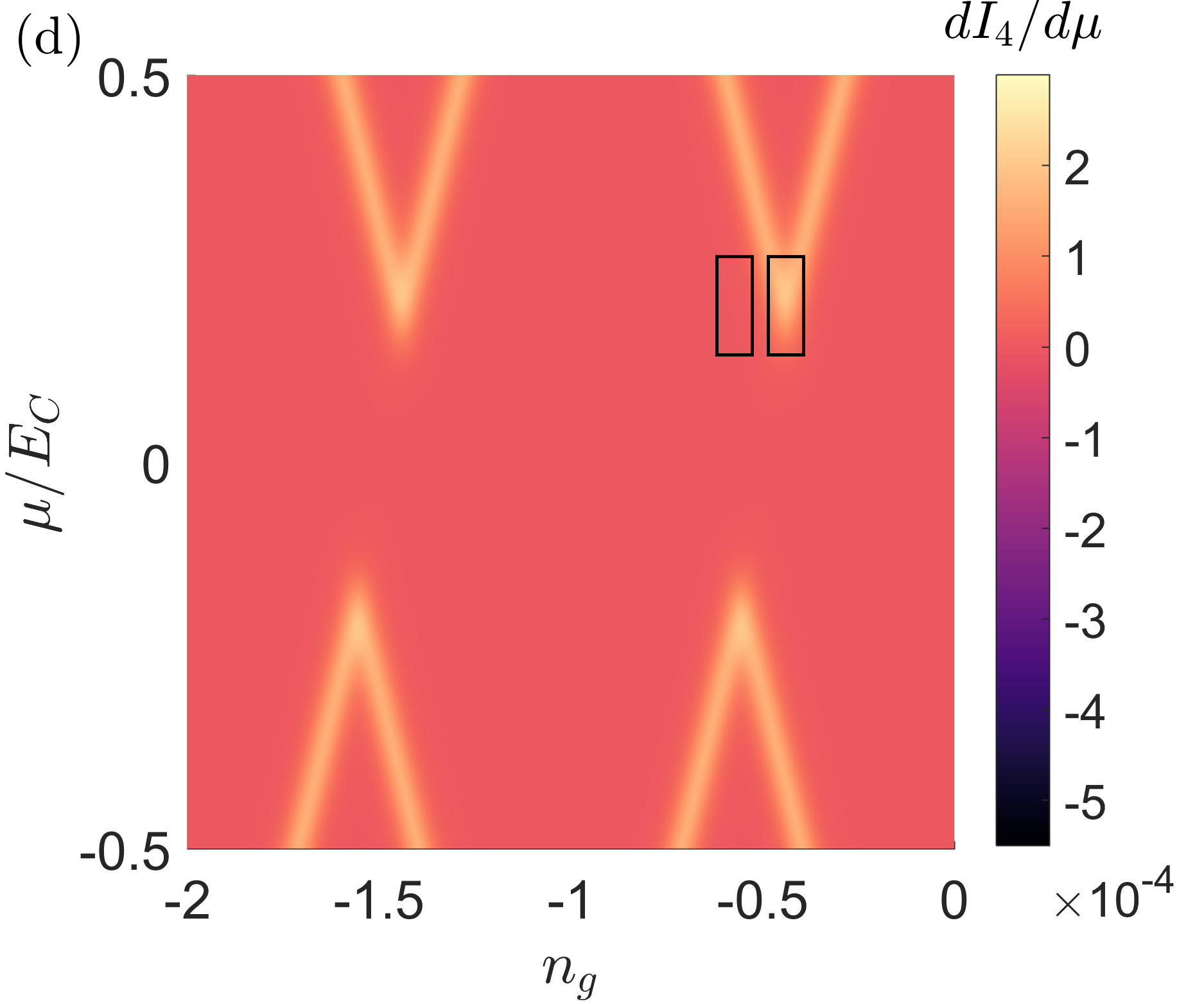}
\caption{Differential conductances as functions of $\mu$ for the Majorana box. A chemical potential $\mu_{1} = \mu$ is applied on lead $1$ whereas identical chemical potential $\mu_{2} = \mu_3 = \mu_4 = -\mu$ are applied on leads $2-4$. (a) $dI_{1}/d\mu$, (b) $dI_{2}/d\mu$, (c) $dI_{3}/d\mu$ and (d) $dI_{4}/d\mu$. In (d), the left box marks a value of $n_{g}$ for which no peak in the differential conductance is observed. The right box marks a value of $n_{g}$ for which a peak in the differential conductance is observed. The parameters are $\xi_{1} = 0.1$, $\xi_{2} = 0.3$, $\Gamma_{i} = 10^{-4}$, $E_{C} = 1$ and $\beta = 25$.}
\label{Fig: Sd_Box}
\end{figure}

We obtain the stability diagrams for the Majorana box by plotting the differential conductances and varying the chemical potential and gate voltage. To be specific, we focus on the current in lead $1$ and apply a symmetric bias such that lead $1$ is held at chemical potential $\mu_{1} = \mu$ while leads $2$--$4$ are held at the same chemical potentials $\mu_2 = \mu_3 = \mu_4 = -\mu$. Varying the gate voltage corresponds to changing $n_g$. The stability diagrams are shown in Fig.~\ref{Fig: Sd_Box}.

The MBSs on both wires are at a finite distance which implies that they are overlapping. For the plot, we choose the overlaps as $\xi_{1} = 0.1$ and $\xi_{2} = 0.3$. For small bias voltages $|\mu|$ transport occurs on the first wire but not on the second one. Transport on the first wire is due to sequential tunneling between the states $\ket{00;N_C}$ and $\ket{10;N_C}$, occurring when the gate voltage is tuned such that two charge states are degenerate. This parameter constellation corresponds to $n_{g}^0 = -(\xi_{1} + E_{C})/2E_{C} = -0.55$.

Increasing the bias leads to a current also in leads $3$ and $4$ [see Fig.~\ref{Fig: Sd_Box}(c,d)]. However, we only observe a conductance peak to the right of $n_g^0$. For instance, if we fix the chemical potential to $\mu = \xi_2 - \xi_1 = 0.2$, we observe a peak for $n_{g} = - ( \xi_{1}- \xi_{2} + 2E_{C})/{4E_{C}} = -0.45 $, whereas no peak is present for $n_g = -0.65 $, see boxes in Fig.~\ref{Fig: Sd_Box}(d). The presence or absence of a peak in the differential conductance is due to the states available for the given chemical potential and gate voltage, which depends on $\xi_{1}$ and $\xi_{2}$. For $n_{g} = -0.65$, only $\ket{00;N_{C}}$ and $\ket{10;N_{C}}$ are occupied, so no current flows in lead 3 or lead 4 because they do not couple to the $d_{1}$ mode. On the contrary, when $n_{g} = -0.45$, the states $\ket{11;N_{C}-1}$, $\ket{00;N_{C}}$, $\ket{10;N_{C}}$ and $\ket{01;N_{C}}$ all have a finite occupation probability, so transport is enabled in all leads.

\begin{figure}[t]
\includegraphics[trim=9.2cm 5.1cm 0.4cm 4.5cm, clip,scale=0.43]{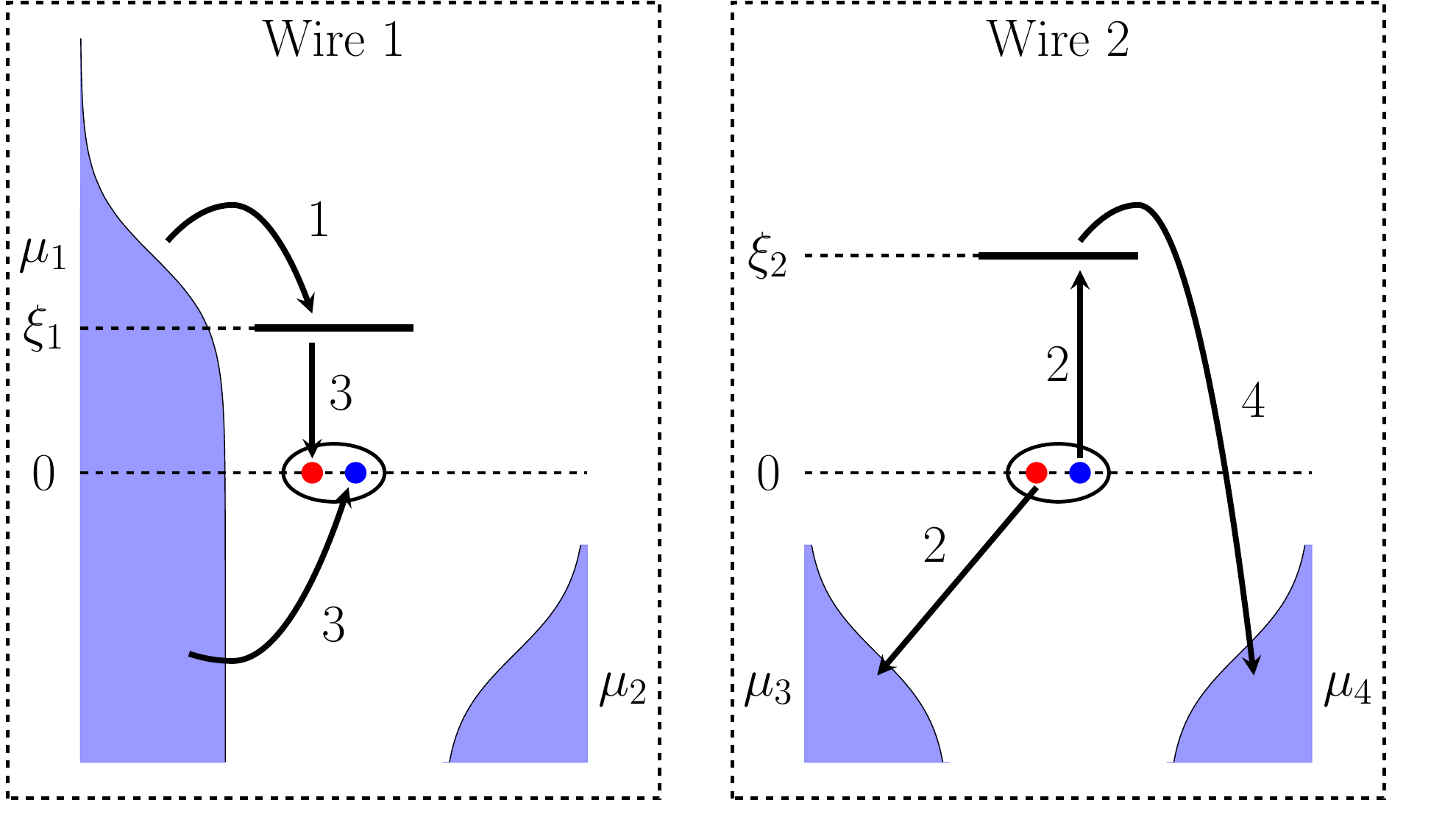}
\caption{Illustration of electron transport from the first to the second wire via the superconductor. Step 1: An electron from lead one tunnels into an MBS on the first wire, $\ket{00;N_C} \to \ket{10;N_C}$. Step 2: A Cooper pair is broken up into one electron which is transferred to an MBS on the second wire, and another one going into lead 3, $\ket{10;N_C} \to \ket{11;N_C-1}$. Step 3: The electron occupying the MBS on the first wire forms a Cooper pair with an electron from lead 1, $\ket{11,N_C-1} \to \ket{01,N_C}$. Step 4: Finally, the electron occupying the MBS on the second wire is transferred to lead 4, $\ket{01,N_C} \to \ket{00,N_C}$. }
\label{Fig: Transport_SeqT}
\end{figure}

To better understand the contributing processes for $n_{g} = -0.45$, consider Fig.~\ref{Fig: Transport_SeqT}. Initially, an electron tunnels from lead 1 into the $d_1$ mode, i.e., it occupies the $\ket{10;N_{C}}$ state. For low enough chemical potential of leads 3 and 4, a Cooper pair can be split into one electron occupying the $d_{2}$ mode, and another one tunneling into lead 3 or 4. The island is then in the $\ket{11;N_C-1}$ state. Next, the electron occupying the $d_{1}$ mode and an electron from, say, lead 2 can form a new Cooper pair. In total, an electron has been transferred from the first wire to the second wire. This is a signature of nonlocal transport between the two wires, mediated by the Cooper pairs.

\begin{figure} [t]
\centering
\includegraphics[trim=6cm 8cm 7cm 8cm, clip,scale=0.45]
{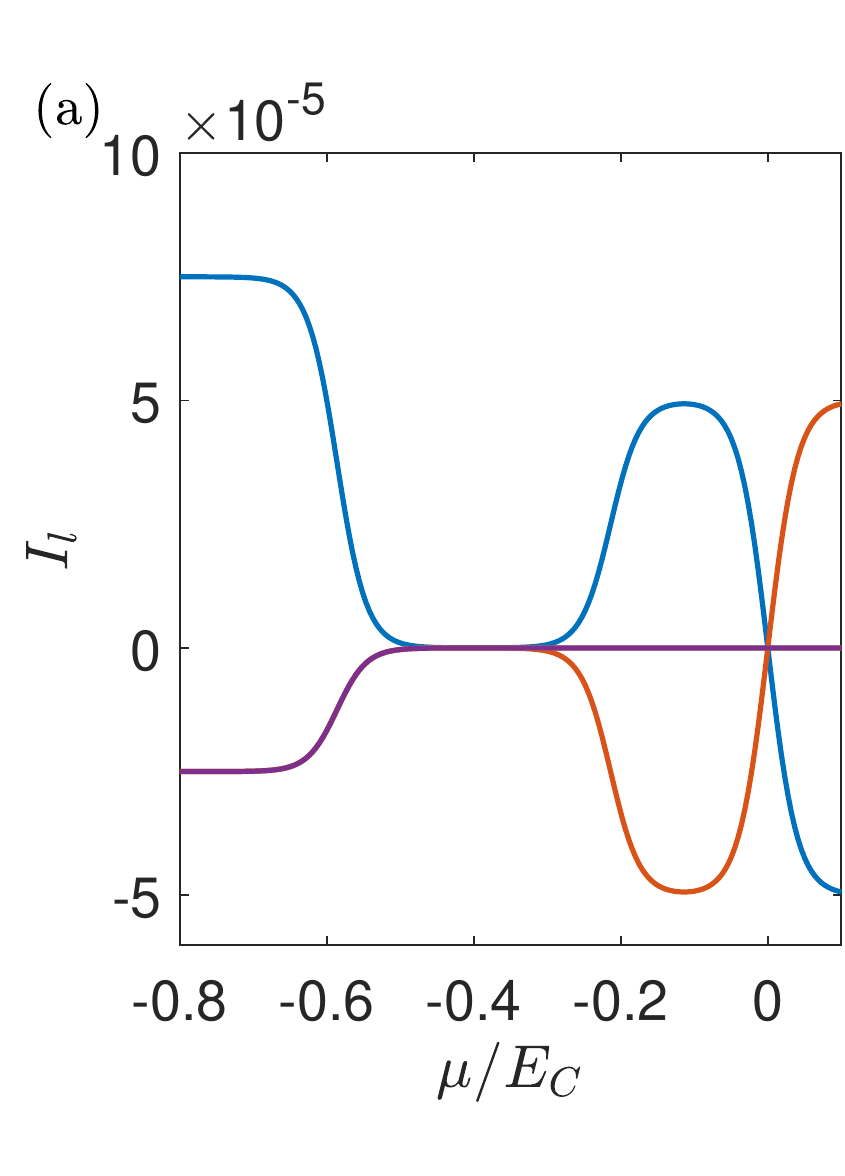}
\includegraphics[trim=5cm 8cm 7cm 8cm, clip,scale=0.45]
{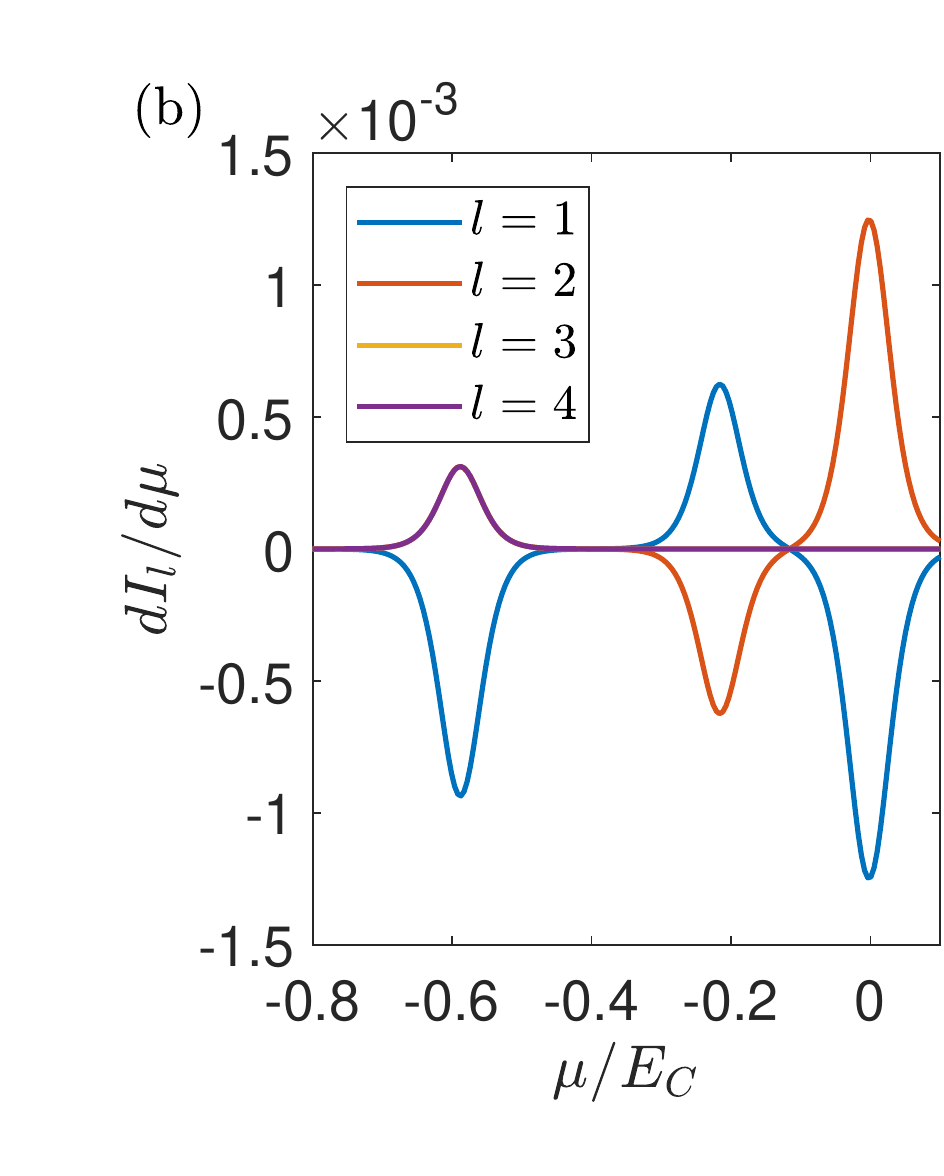}
\caption{(a) Currents and (b) differential conductances as functions of $\mu$ for the Majorana box. A chemical potential $\mu_{1} = \mu$ is applied on lead $1$ whereas identical chemical potential $\mu_{2} = \mu_3 = \mu_4 = -\mu$ are applied on leads $2-4$. The parameters are $\xi_{1} = 0.3$, $\xi_{2} = 0.5$, $\Gamma_{i} = 10^{-4}$, $E_{C} = 1$ and $\beta = 50$. Nonlocal Coulomb blockade causes all currents to vanish in the bias region $-0.6 < \mu/E_C < -0.2$.}
\label{Fig: IdI_Mbox}
\end{figure}

In Fig.~\ref{Fig: Sd_Box}(a) we observe, in certain regions of bias and gate voltage, a positive differential conductance, in stark contrast to the negative differential conductance observed throughout the rest of the stability diagram. To highlight this feature, let us consider the parameters $\xi_{1} = 0.3$ and $\xi_{2} = 0.5$, and the system is tuned on resonance for the first wire, i.e., $n_{g} = n_g^0$. The current and the differential conductance are plotted in Fig.~\ref{Fig: IdI_Mbox}. One finds that the current as a function of bias voltage is non-monotonic, and even vanishes in a range of negative bias voltages. This can be regarded as a nonlocal Coulomb blockade phenomenon and can be explained as follows: for bias voltages $\mu_{3,4} > \xi_{2} - \xi_{1}$, the state $\ket{01;N_{C}}$ will be occupied. Hence, for transport to occur in the first wire the system would need to make a transition from the $\ket{01;N_{C}}$ to the $\ket{11;N_{C}}$ state. However, this transition is impossible because the state $\ket{11;N_{C}}$ has a larger charging energy.

\subsubsection{T-junction}

\begin{figure}[t]
\includegraphics[scale=0.26]{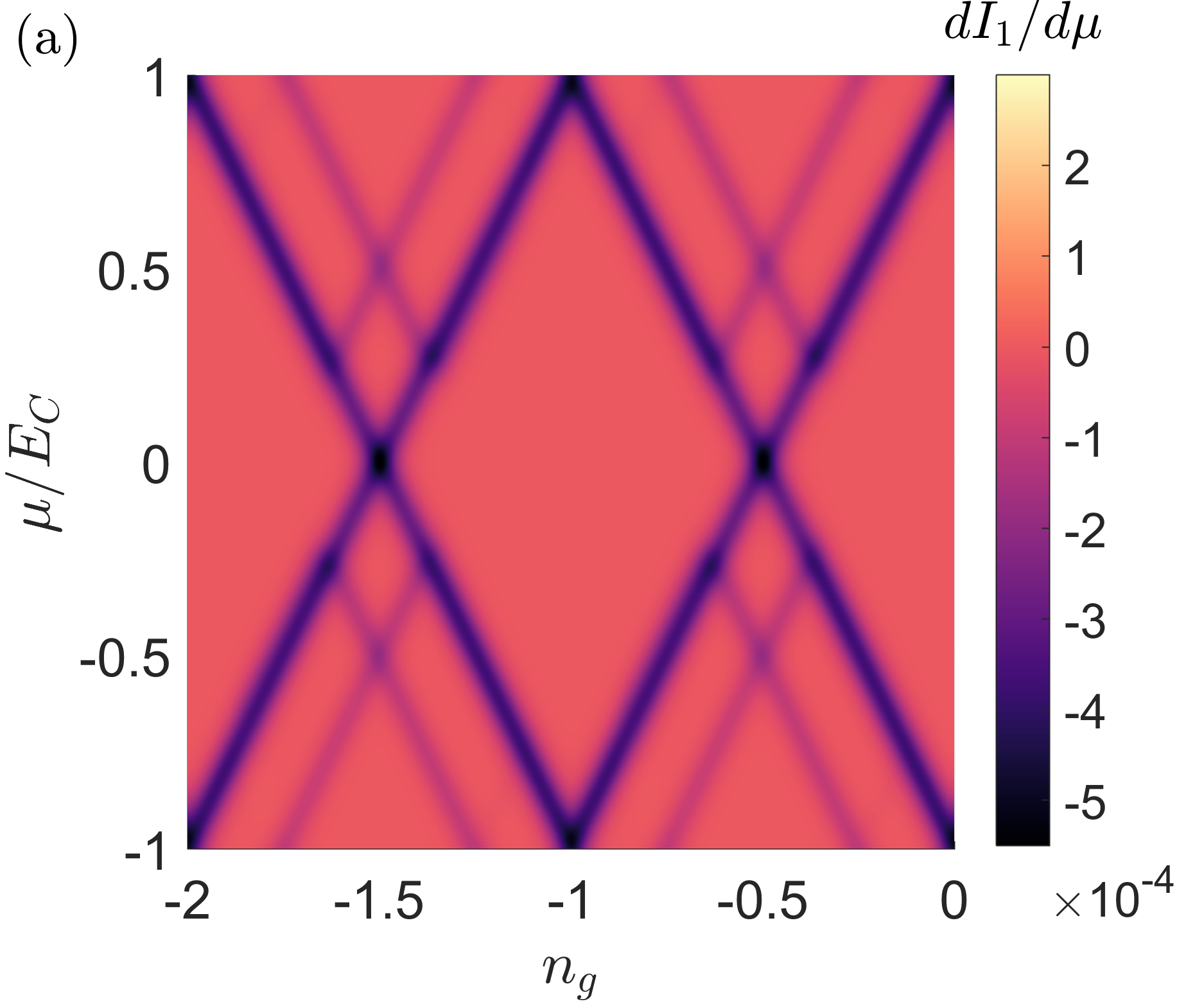}
\includegraphics[scale=0.26]{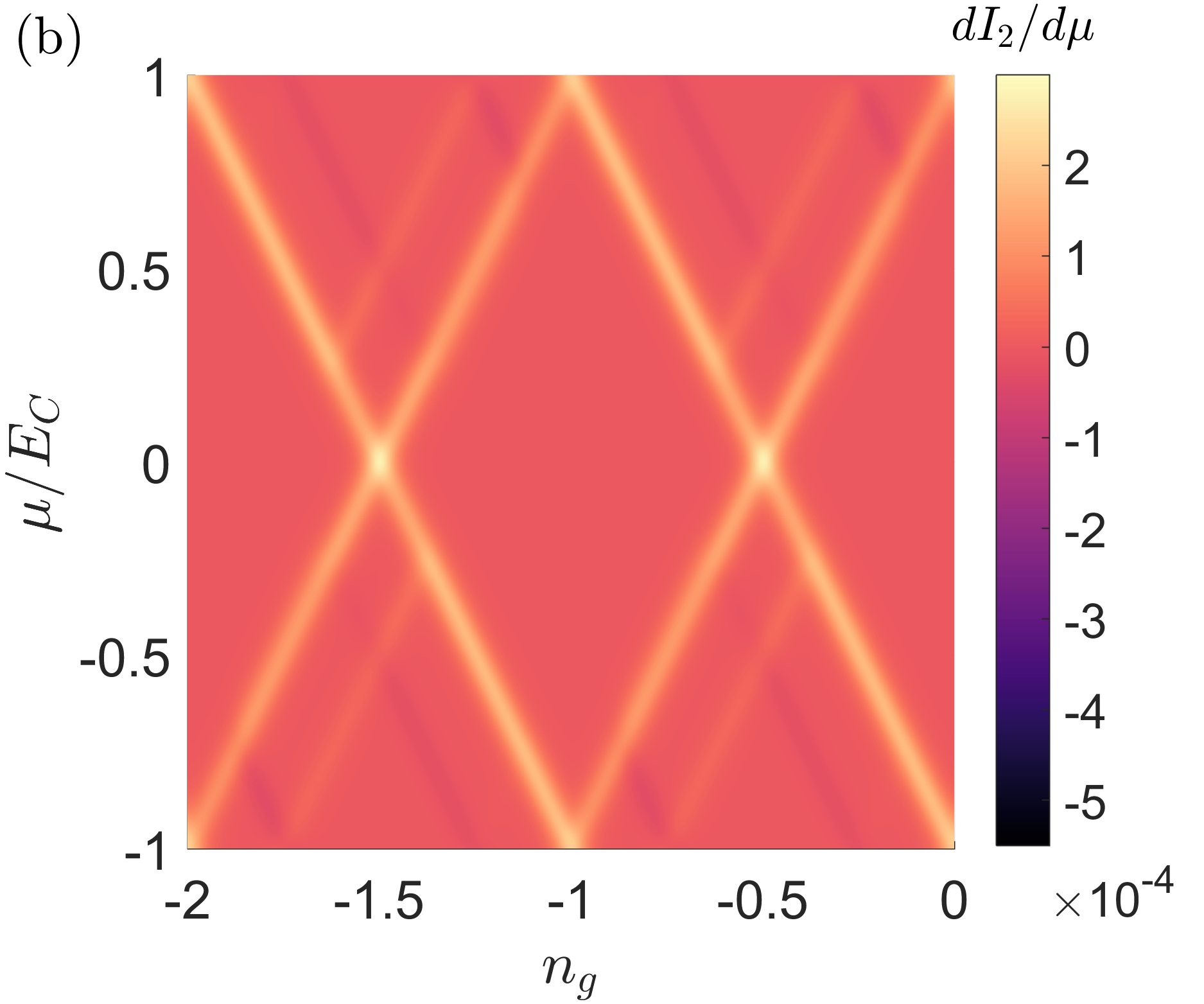}\\
\includegraphics[scale=0.26]{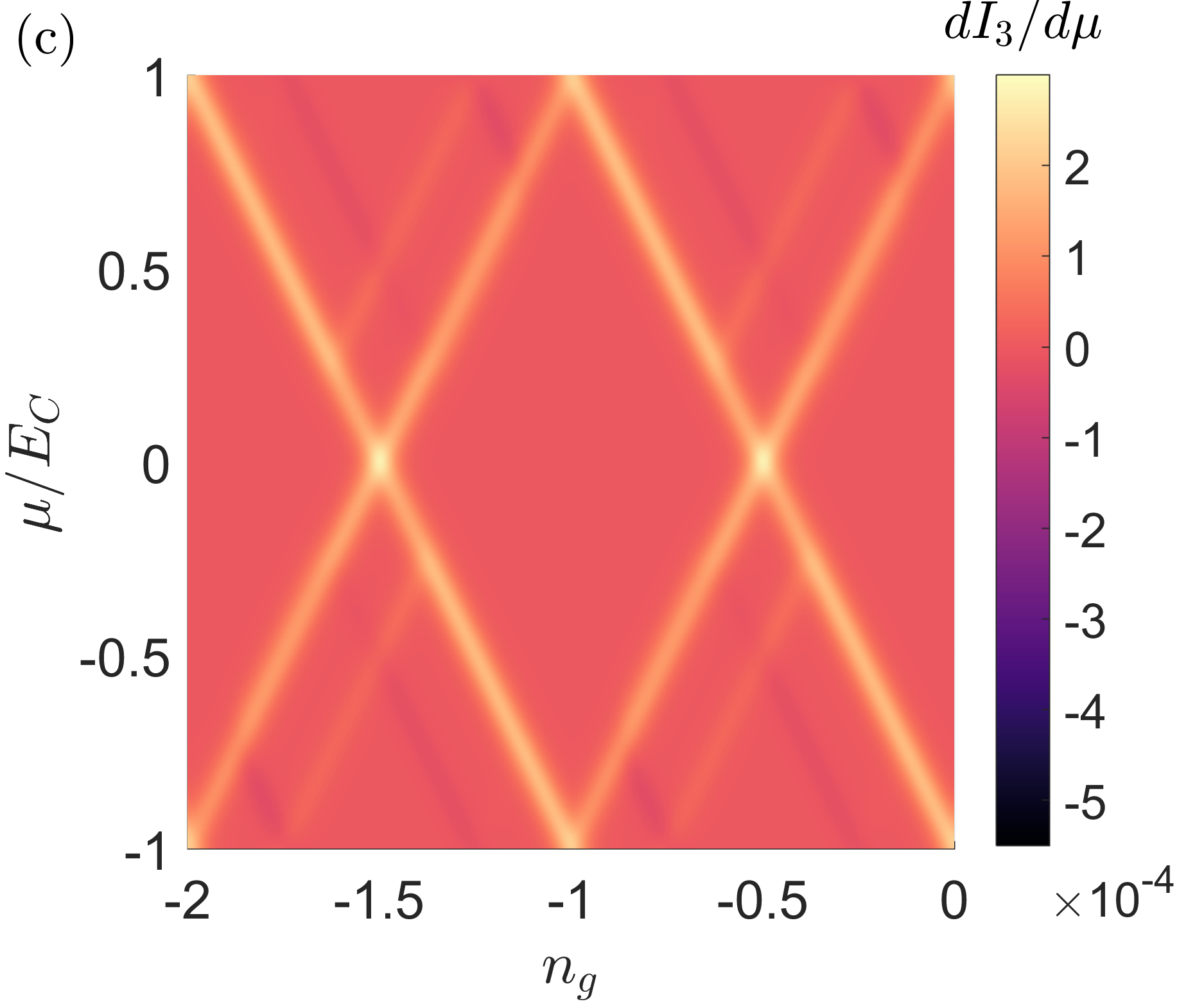}
\includegraphics[scale=0.26]{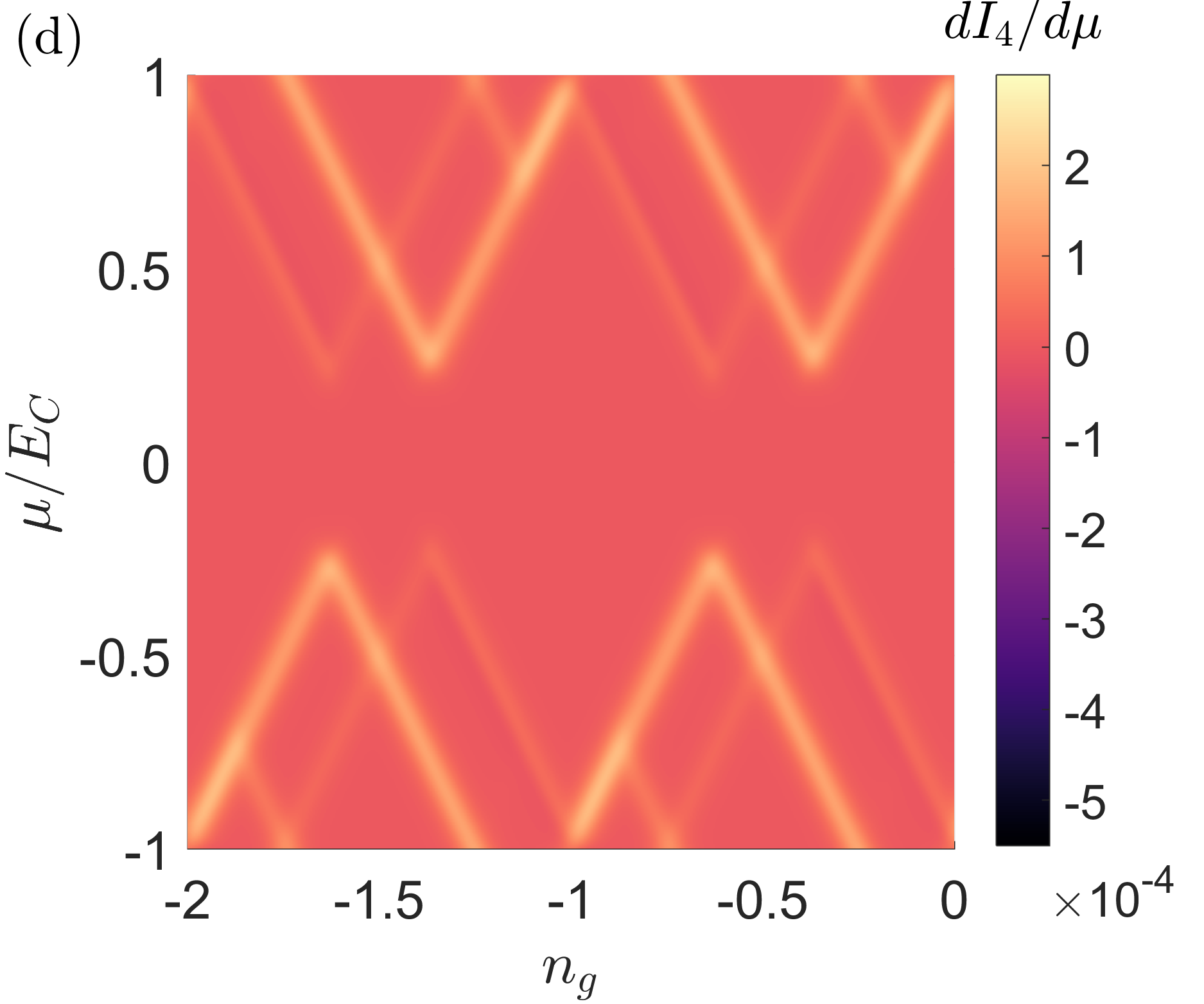}
\caption{Differential conductances as functions of bias voltage $\mu$ and gate voltage $n_g$ for the T-junction at a symmetric bias configuration. Here, the chemical potential $\mu_{1} = \mu$ is applied to lead $1$, whereas the other leads have equal chemical potentials $\mu_{2} = \mu_3 = \mu_4 = -\mu$. The subplots show (a) $dI_{1}/d\mu$, (b) $dI_{2}/d\mu$, (c) $dI_{3}/d\mu$ and (d) $dI_{4}/d\mu$. The parameters are $\xi_{T} = 0.5$, $\Gamma_{l} = 10^{-4}$, $E_{C} = 1$ and $\beta = 25$.}
\label{Fig: Sd_T}
\end{figure}

Next we investigate the sequential tunneling regime for the T-junction. As a point of reference, let us briefly review the noninteracting limit $E_C = 0$. Assuming that $|\mu_{l}|, T \ll \xi_{T}$, the current is
\begin{equation}
\langle I_{l} \rangle = \Gamma_{l}|\alpha_{l1}|^{2}\left[ 1 - 2n_{F}(-\mu_{l})\right].
\end{equation}
Up to linear order in $\Gamma_l$, this reproduces the result found in Ref.~\cite{weithofer_electron_2014}, where an exact solution for the noninteracting limit was derived. Moreover, we confirm that up to order $\Gamma_l$, no current flows in lead $4$ because $\alpha_{41} = 0$, see Table~(\ref{Tab: Tjunction}). The leading-order process at the central lead would be a double crossed Andreev process of higher order in $\Gamma_l$ \cite{weithofer_electron_2014}.

\begin{figure}
\includegraphics[trim=9.4cm 8.5cm 6cm 4cm, clip,scale=0.53]{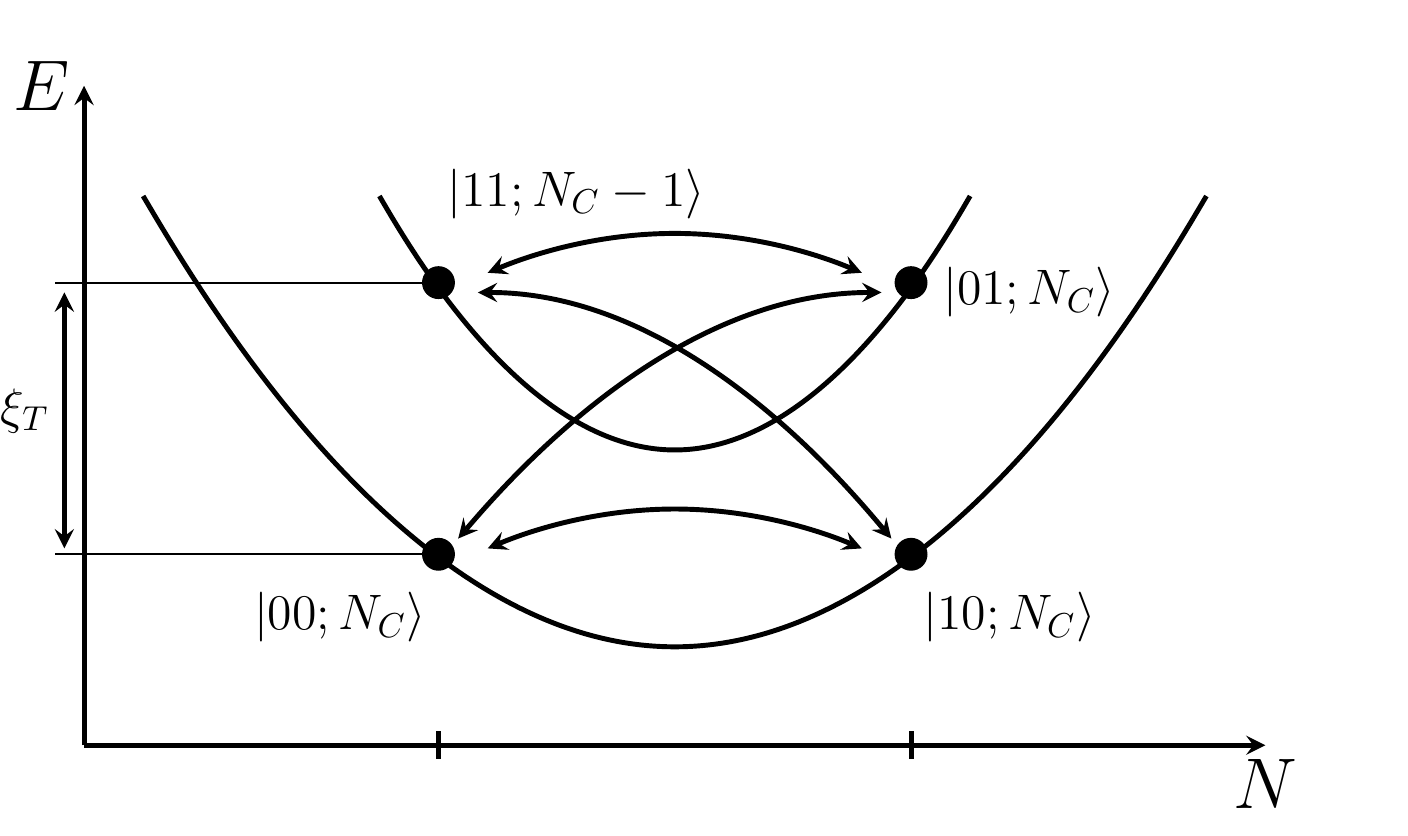}
\caption{Charge states contributing to sequential tunneling for $n_{g} = -1/2$ in the T-junction. For small applied biases the only allowed transitions are between $\ket{00;0} \rightleftharpoons \ket{10;0}$. Increasing the bias window above the energies of the lowest excited states allows for the island to cycle between the states $\ket{00;N_{C}} \rightleftharpoons \ket{10;N_{C}}$, $\ket{10;N_{C}} \rightleftharpoons \ket{11;N_{C}-1}$, $\ket{01;N_{C}} \leftrightharpoons \ket{11;N_C-1}$ and $\ket{00;N_{C}}\leftrightharpoons \ket{01;N_{C}}$.}
\label{Fig: ParabolaZoom_SeqT}
\end{figure}

For a nonzero charging energy $E_C$, we obtain the stability diagrams, see Fig.~\ref{Fig: Sd_T}. As for the Majorana box we apply a symmetric bias such that lead $1$ is held at chemical potential $\mu_{1} = \mu$ while leads $2$--$4$ are held at the same chemical potentials $\mu_2 = \mu_3 = \mu_4 = -\mu$. At the degeneracy points $n_{g} = -1/2$, a zero-bias conductance peak is observed in the outer leads, whereas transport on the central lead is blocked. In this parameter range, sequential tunneling of electrons is possible due to switching between the island states $\ket{00;N_{C}} \rightleftharpoons \ket{10;N_{C}}$, as depicted in Fig.~\ref{Fig: ParabolaZoom_SeqT}.

Increasing the bias window to values $|\mu| > \xi_T$ allows for a current in the central lead as well and provides additional transport processes at the outer leads, visible as sidebands in Fig.~\ref{Fig: Sd_T}. The transport between the different leads is carried both by single electrons being transported through the MBSs as well as by a process involving the additional creation and annihilation of Cooper pairs. The island state correspondingly changes between the states $\ket{00;N_{C}} \rightleftharpoons \ket{10;N_{C}}$, $\ket{10;N_{C}} \rightleftharpoons \ket{11;N_{C}-1}$, $\ket{01;N_{C}} \leftrightharpoons \ket{11;N_C-1}$ and $\ket{00;N_{C}}\leftrightharpoons \ket{01;N_{C}}$, as seen in Fig.~\ref{Fig: ParabolaZoom_SeqT}.

\subsubsection{Nonlocal transport and teleportation} \label{Sec: Non-local}
The nonlocal transport between the two wires of the Majorana box (see Fig.~\ref{Fig: Sd_Box}) can be interpreted along the lines of the ''teleportation'' processes found in Ref.~\cite{fu_electron_2010}. In this context, electron teleportation is defined as a nonlocal current which is independent of the length of the wires, and thus independent of the overlap of the MBSs. Hence, to isolate the teleportation contribution, let us briefly discuss structures where the wires are infinitely long, corresponding to the limit where $\xi_1 = \xi_2 = \xi_T = 0$. In this limit, the Majorana box and T-junction structures we investigate become indistinguishable from each other.

\begin{figure} [t]
\includegraphics[trim=4cm 10cm 5cm 9cm, clip,scale=0.65]{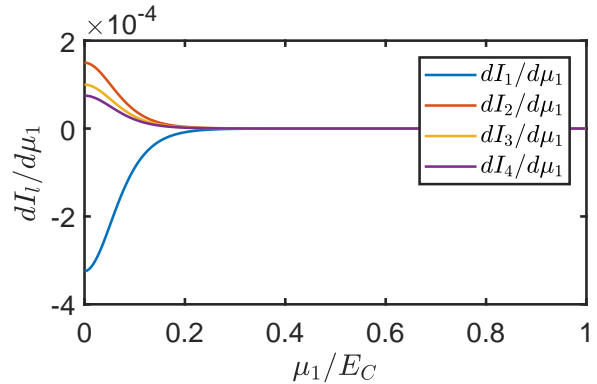}
\caption{The nonlocal conductance $dI_{l}/d\mu_{1}$ in a system with four MBSs. Note that $dI_{2}/d\mu_{1} = dI_{3}/d\mu_{1} = dI_{4}/d\mu_{1}$. A bias voltage $\mu_{1}$ is applied on lead $1$, whereas the other leads are grounded. The overlap between the MBSs is zero. The other parameters are $\Gamma_{1} = 10^{-4}$, $\Gamma_{2} = \Gamma_{1}/2$, $\Gamma_{3} = \Gamma_{1}/3$, $\Gamma_{4} = \Gamma_{1}/4$, $E_{C} = 1$ and $\beta = 25$.}
\label{Fig: non-local}
\end{figure}

We apply a bias voltage $\mu_{1}$ to lead $1$, and assume all other leads to be grounded. This allows us to calculate the nonlocal differential conductance $dI_{l}/d\mu_{1}$ for $l \in \{1, \ldots, 4\}$.  The results are plotted in Fig~\ref{Fig: non-local}. A zero-bias differential conductance peak is observed in all leads. This is a signature of nonlocal currents in leads 2-4 due to an applied bias on lead 1. These nonlocal currents can be interpreted as electron teleportation \cite{fu_electron_2010}; electrons tunnel via the MBSs independent of the length of the wire.

\subsection{Cotunneling} \label{Sec: CoT}

\begin{figure}
\includegraphics[trim=9.4cm 8.5cm 4cm 4cm, clip,scale=0.53]{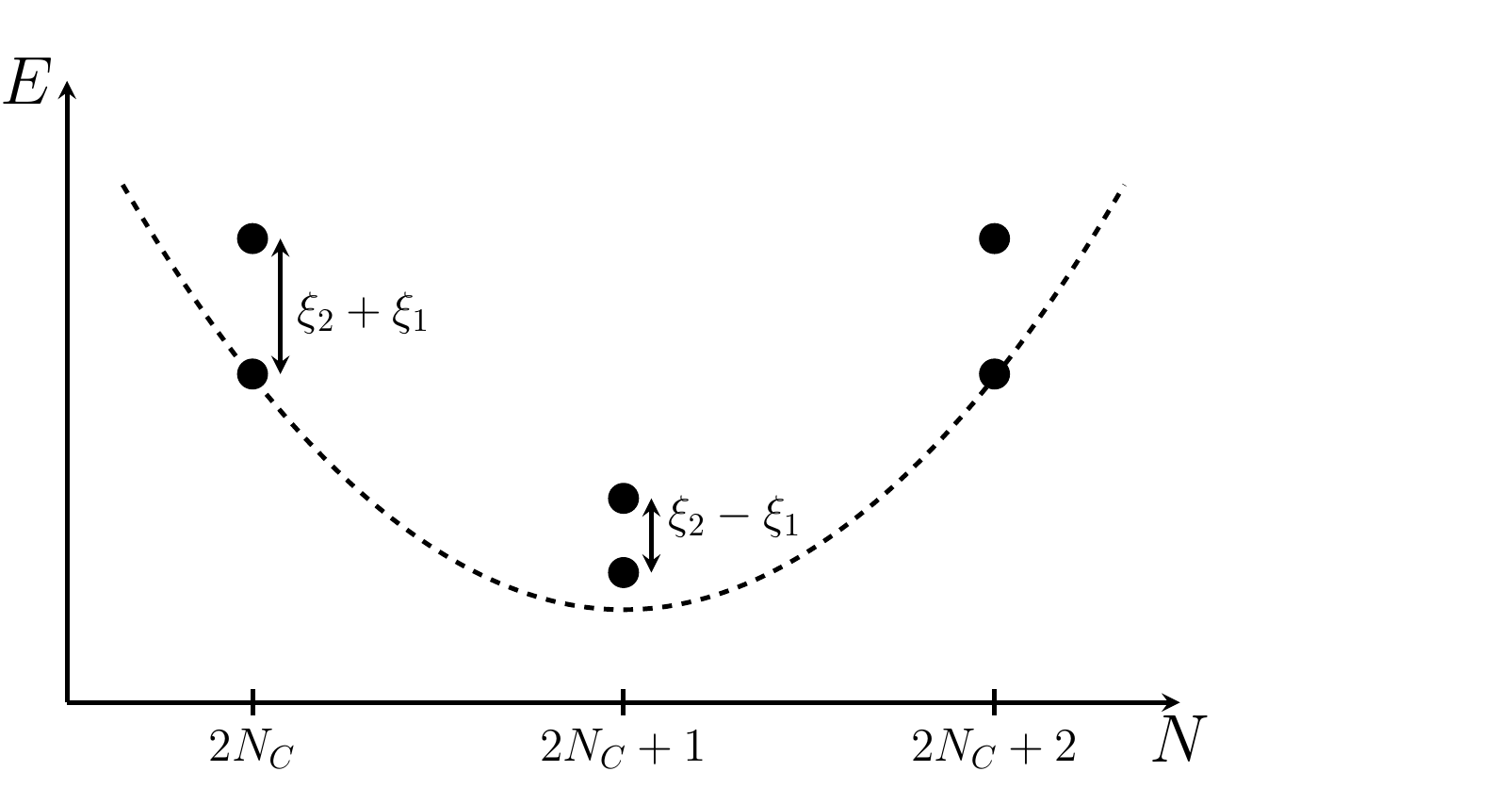}
\caption{Typical charge state distribution in the cotunneling regime. The plot shows the energies of the six charge states with the lowest energies for $n_{g} = -1$. For this figure we have assumed $\xi_{2} > \xi_{1}$.}
\label{Fig: ParabolaZoom}
\end{figure}

In the cotunneling regime, see Fig.~\ref{fig: parabolas}(b), the island ground state has no degeneracies between different charge states, so transport can only occur via the virtual occupation of an intermediate state at higher energies. The starting point for the calculation is the rate equation for the occupation probabilities. In general we have, schematically,
\begin{equation}
\dot{P}_{\alpha} = -\sum_{\alpha}W_{\beta}^{\alpha}P_{\alpha} + \sum_{\beta} W^{\beta}_{\alpha}P_{\beta}.
\label{Eq: Occ1}
\end{equation}
The equation describes the rate of change in the occupation probability $P_\alpha$ of the island state $\ket{\alpha} = \ket{\mathbf{n}; N_C}$. The first term on the right hand side describes the decrease of occupancy of this state, while the second term describes its increase. The parameters $W^{\alpha}_{\beta}$ denote the transition rates between a given initial state $\ket{\alpha}$ and a final state $\ket{\beta}$. Assuming that $|\mu_l| \ll E_{C}$, it is sufficient to take into account the six charge states lowest in energy, see Fig.~\ref{Fig: ParabolaZoom}. Since either $d_{1}$ or $d_{2}$ can be occupied, there are two possibilities for each charge state. For instance, for $n_g = -1$, the states under consideration are $\ket{11;N_{C}-1}$, $\ket{00;N_{C}}$, $\ket{10;N_{C}}$, $\ket{01;N_{C}}$, $\ket{00;N_{C} + 1}$ and $\ket{11;N_{C}}$. As seen in Fig.~\ref{Fig: ParabolaZoom}, the low energy states are $\ket{10;N_{C}}$ and $\ket{01;N_{C}}$. The virtual states are $\ket{00;N_{C}}$, $\ket{11;N_{C}-1}$, $\ket{00;N_{C}+1}$ and $\ket{11;N_{C}}$. For simplicity, we will use $N_{C} = 0$ in the following. As before, this can be done without any loss of generality because of the periodicity of the conductance as a function of the charge on the island. The occupation probabilities for the states are then denoted by $P_{11;-1}$, $P_{00;0}$, $P_{10;0}$, $P_{01;0}$, $P_{00;1}$ and $P_{11;0}$. We require that $\sum_{\alpha}P_{\alpha} = 1$.

The transition rates can be calculated from Fermi's golden rule in the $T$ matrix representation,
\begin{equation}
W^{\alpha}_{\beta} = 2\pi\sum_{i,f}\left| \bra{\psi_f, \beta} \hat{T} \ket{\psi_i, \alpha} \right|^{2}\delta(E_{f,\beta} - E_{i,\alpha}),
\label{Eq: rates}
\end{equation}
where $\ket{\psi_{i,f}}$ denotes the initial (final) state of the leads, and $E_{i,\alpha}$ and $E_{f,\beta}$ are the initial and final energy, respectively, for a transition from initial state $\ket{\psi_i, \alpha}$ to final state $\ket{\psi_f,\beta}$. The $T$ matrix reads
\begin{equation}
\hat{T} = H'_{\rm tun} + H'_{\rm tun}\frac{1}{H_{0} - E_{i,\alpha}}\hat{T},
\label{Eq: Tmatrix}
\end{equation}
where the unperturbed Hamiltonian is given by
\begin{equation}
H_{0} = H_{\rm leads} + H_{\rm MBS} + H_{\rm charging}.
\label{Eq: H0}
\end{equation}
Expanding Eq.~(\ref{Eq: Tmatrix}) to second order in $H_{\rm tun}^{\prime}$, we obtain the second order transition rates from Eq.~(\ref{Eq: rates}). The full expressions are given in App.~\ref{Apn: rates}.

By considering that in the steady state $\dot{P}_{\alpha} = 0$, Eq.~(\ref{Eq: Occ1}) can be written as $\underline{W} \mathbf{P} = \mathbf{0} $, where $\underline{W}$ is a matrix containing the transition rates $W^{\alpha}_{\beta}$ and $\mathbf{P} = (P_{11;-1}, P_{00;0}, P_{10;0},P_{01;0},P_{00;1},P_{11;0})$. We solve for $\mathbf{P}$ numerically and obtain, for $n_{g} \approx -1$ and when the applied bias is smaller than $E_{C}$, that $P_{11;-1} \approx P_{00;0} \approx P_{00;1} \approx P_{11;0} \propto e^{-E_{C}/T} \approx 0$, whereas $P_{10;0}$ and $P_{01;0}$ remain finite and depend on the bias configuration. This is to be expected since if we do not apply a sufficient bias there is not enough energy for electrons to perform a real transition from the two lowest states to the states higher in energy. Still, a current is possible by transitions via virtual states.

As in Eq.~(\ref{Eq: CurrentSQ}) the current in lead $l$ is given by
\begin{align}
\langle I_{l} \rangle &= \sum_{\mathbf{n},N_{C}}\left( W_{l}^{1,+} - W_{l}^{1,-} \right) P(\mathbf{n},N_{C}) \notag \\
&+ 2\sum_{\mathbf{n},N_{C}}\left( W_{l}^{2,+} - W_{l}^{2,-} \right) P(\mathbf{n},N_{C}),
\label{Eq: Current2}
\end{align}
where $W_{l}^{1,+(-)}$ are the transition rates for those processes, in Eq.~(\ref{Eq: rates}), which increase (decrease) the number of particles in lead $l$ by one, whereas $W_{l}^{2,+(-)}$ correspond to processes that increase (decrease) the number of particles in lead $l$ by two. Since only $P_{10;0}$ and $P_{01;0}$ are non-zero, the only contributions to the current are proportional to $W_{10;0}^{10;0}$, $W_{01;0}^{01;0}$, $W_{01;0}^{10;0}$ and $W_{10;0}^{01;0}$. This means that the current is due to transitions where either the initial and final state of the island are identical (elastic cotunneling, ECT) or where the initial and final state of the island are different (inelastic cotunneling, ICT). In both cases the total charge on the island is conserved.

The expression for the current, Eq.~(\ref{Eq: Current2}) can now be reduced to
\begin{equation}
\langle I_{l} \rangle = I^{\rm ECT}_{l} + I^{\rm ICT}_{l},
\label{Eq: CTcurrent}
\end{equation}
where
\begin{align}
I^{\rm ECT}_{l} &= \sum_{k}\Big\lbrace\left[W_{10;0}^{10;0}(l,k) - W_{10;0}^{10;0}(k,l)\right]P_{10;0} \nonumber \\ & \hspace{26pt}+ \left[W_{01;0}^{01;0}(l,k) - W_{01;0}^{01;0}(k,l)\right]P_{01;0}\Big\rbrace,
\end{align}
and
\begin{align}
I^{\rm ICT}_{l} &= \sum_{k}\Big\lbrace \left[ W_{10;0}^{01;0}(l,k) - W_{10;0}^{01;0}(k,l)\right]P_{01;0} \nonumber \\ &\hspace{26pt}+ \left[W_{01;0}^{10;0}(l,k) - W_{01;0}^{10;0}(k,l)\right]P_{10;0}\Big\rbrace.
\end{align}
Here, $W^{\alpha}_{\beta}(l,k)$ denotes the second-order transition rate from the initial island state $\ket{\alpha}$ to final state $\ket{\beta}$ due to electron tunneling between leads $l$ and $k$.

In the limit $\xi_j, |\mu|\ll E_C$ and $T=0$ we find analytical expressions for these rates. For the Majorana box they are given by
\begin{align}
W^{10;0}_{10;0}(l,k) &= \frac{2W_{l}W_{k}}{\pi E_{C}^{2}} (\mu_{l} - \mu_{k} ) \theta(\mu_{l}- \mu_{k}), \label{Eq: Mbox_ECT}\\
W^{10;0}_{01;0}(l,k) &= \frac{2W_{l}W_{k}}{\pi E_{C}^{2}} (\mu_{l} - \mu_{k} +\xi_1 - \xi_2) \nonumber \\ &\times\theta(\mu_{l} - \mu_{k} +\xi_1 - \xi_2),
\label{Eq: Mbox_ICT}
\end{align}
where $k,l \in \lbrace 1,2 \rbrace$ in the first line, and $k,l\in\lbrace 3,4\rbrace$ in the second line. For the calculations we have assumed a constant density of states, $D_{l} = 1/(2 \pi v_F)$, in the leads and defined $W_{l} = 2\pi |t_{l}|^{2}D_{l}$.  For the T-junction we obtain
\begin{align}
W^{10;0}_{10;0}(l,k) &= \frac{2W_{l}W_{k}}{\pi E_{C}^{2}} (\mu_{l} - \mu_{k} ) \theta(\mu_{l}- \mu_{k}),
\label{Eq: T_ECT} \\
W^{10;0}_{01;0}(l,k) &= \frac{2W_{l}W_{k}}{\pi E_{C}^{2}} (\mu_{l} - \mu_{k} - \xi_T)\theta(\mu_{l} - \mu_{k} - \xi_T),
\label{Eq: T_ICT}
\end{align}
with $k,l\in\{ 1, \ldots, 4\}$. We note that if there is no overlap between the MBSs of the systems, $W^{10;0}_{10;0}(l,k) = W^{01;0}_{01;0}(l,k) = W^{10;0}_{01;0}(l,k) = W^{01;0}_{10;0}(l,k)$ with $k,l\in\lbrace 1,2,3,4\rbrace$. Thus, if the overlap is zero, the strengths of elastic and inelastic cotunneling become equal to each other.

\subsubsection{Majorana box}
\begin{figure}
\includegraphics[trim=5cm 7.5cm 6cm 6cm, clip,scale=0.4]{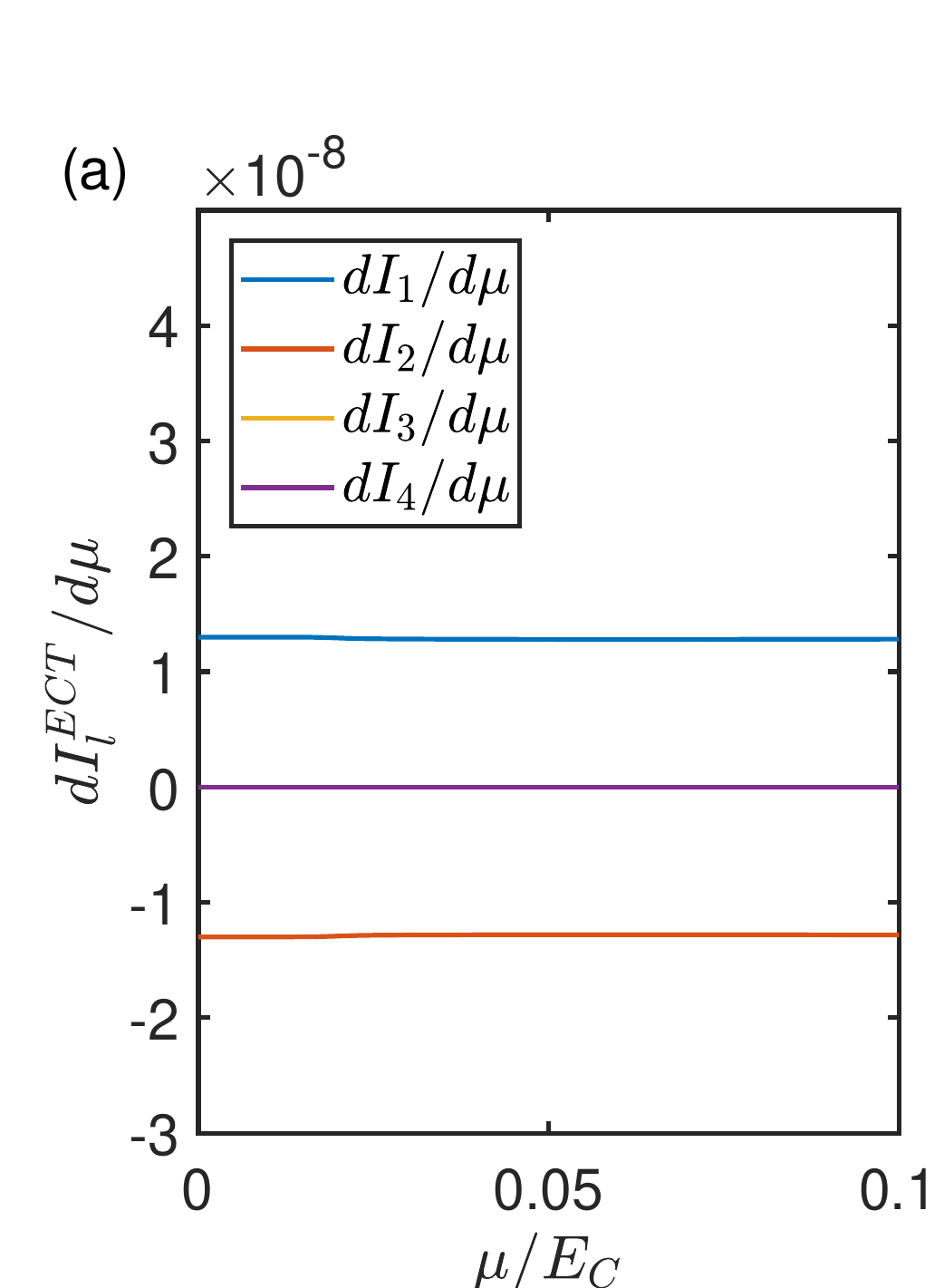}
\includegraphics[trim=5cm 7.5cm 6cm 6cm, clip,scale=0.4]{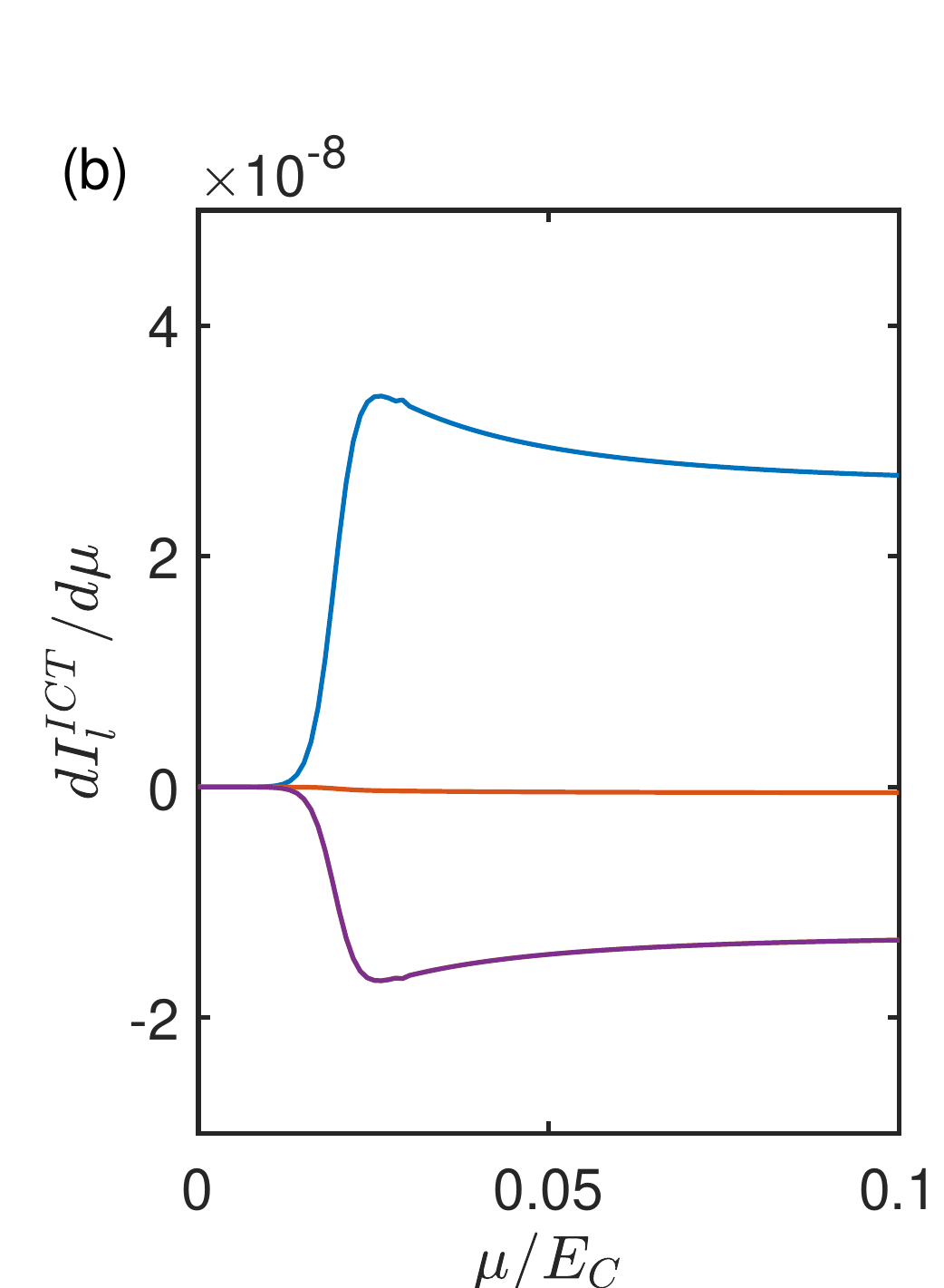}
\caption{ Differential conductance in the cotunneling limit as a function of $\mu$, for the Majorana box setup. A bias $\mu_{1} = \mu$ is applied on lead $1$, whereas a bias $\mu_{2} = -\mu$ is applied on leads $2-4$. (a) Differential conductance due to elastic cotunneling and (b) differential conductance due to inelastic cotunneling. Note that in both plots $dI_{3}/d\mu = I_{4}/d\mu$. The parameters are $E_{C} = 1$, $W_{i} = 1\cdot 10^{-4}$, $\xi_{1} = 0.01$, $\xi_{2} = 0.03$, $\beta = 900$ and $n_{g} = - 1$.}
\label{Fig: CT1}
\end{figure}

We proceed by calculating the differential conductance of the Majorana box. For the plots, we assume again a symmetric bias configuration where $\mu_{1} = \mu$ and $\mu_{2} = \mu_3 = \mu_4 = -\mu$. The results are plotted in Fig.~\ref{Fig: CT1}. We observe a current between leads 1 and 2 due to elastic cotunneling. Comparing the analytical results (\ref{Eq: Mbox_ECT}) with the numerical ones, we note that these are in close agreement. The numerical results show a constant differential conductance which is also obtained analytically (differentiating the transition rates with respect to the bias gives a constant). Since the $d_{1}$ mode does not couple to lead 3 or lead 4, it is not possible to have any current due to elastic cotunneling between lead 1 and the leads 3 and 4.

When the bias is larger than the difference between the overlaps of the MBSs, a current in leads 3 and 4 is observed. This is solely due to inelastic cotunneling. This is also seen in the analytical results (\ref{Eq: Mbox_ICT}), which explains why there is no current when $\mu <|\xi_{2} - \xi_{1}|$.

\begin{figure}
\includegraphics[trim=9.4cm 10cm 2.7cm 4cm, clip,scale=0.48]{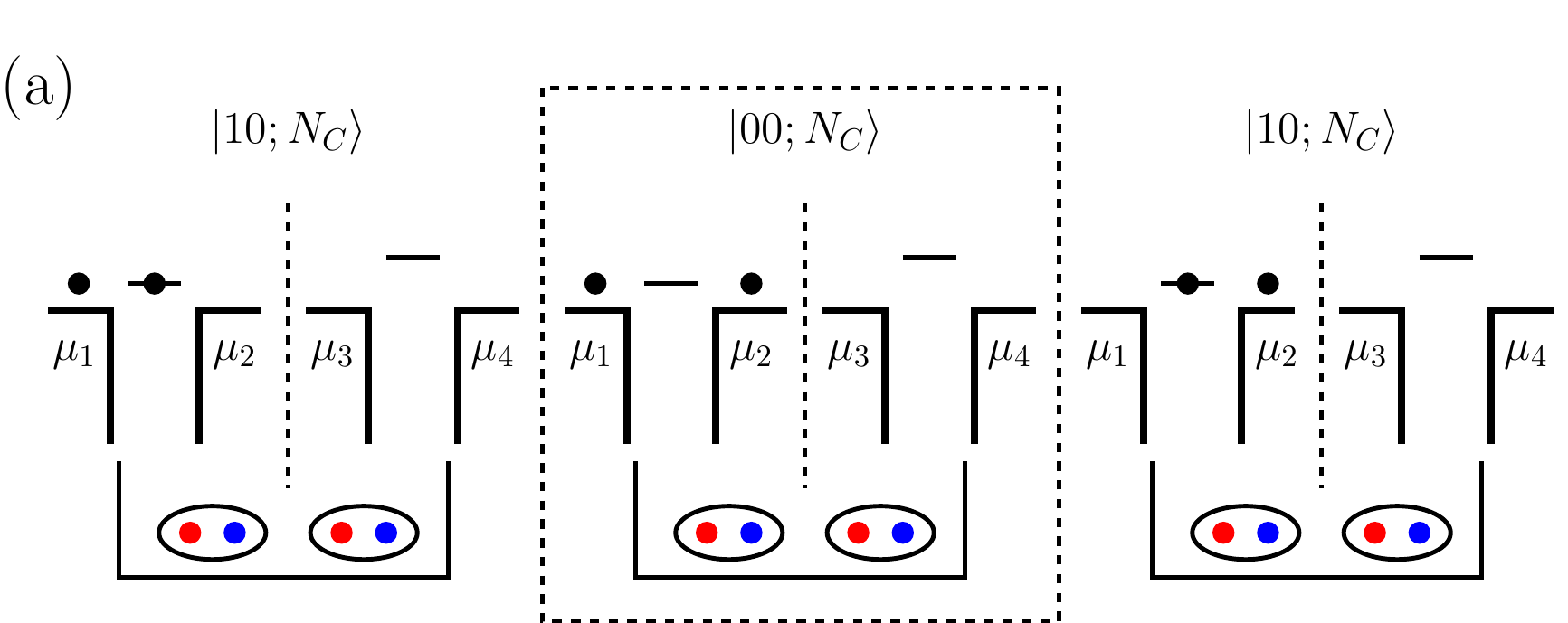}
\includegraphics[trim=9.4cm 10cm 2.7cm 4cm, clip,scale=0.48]{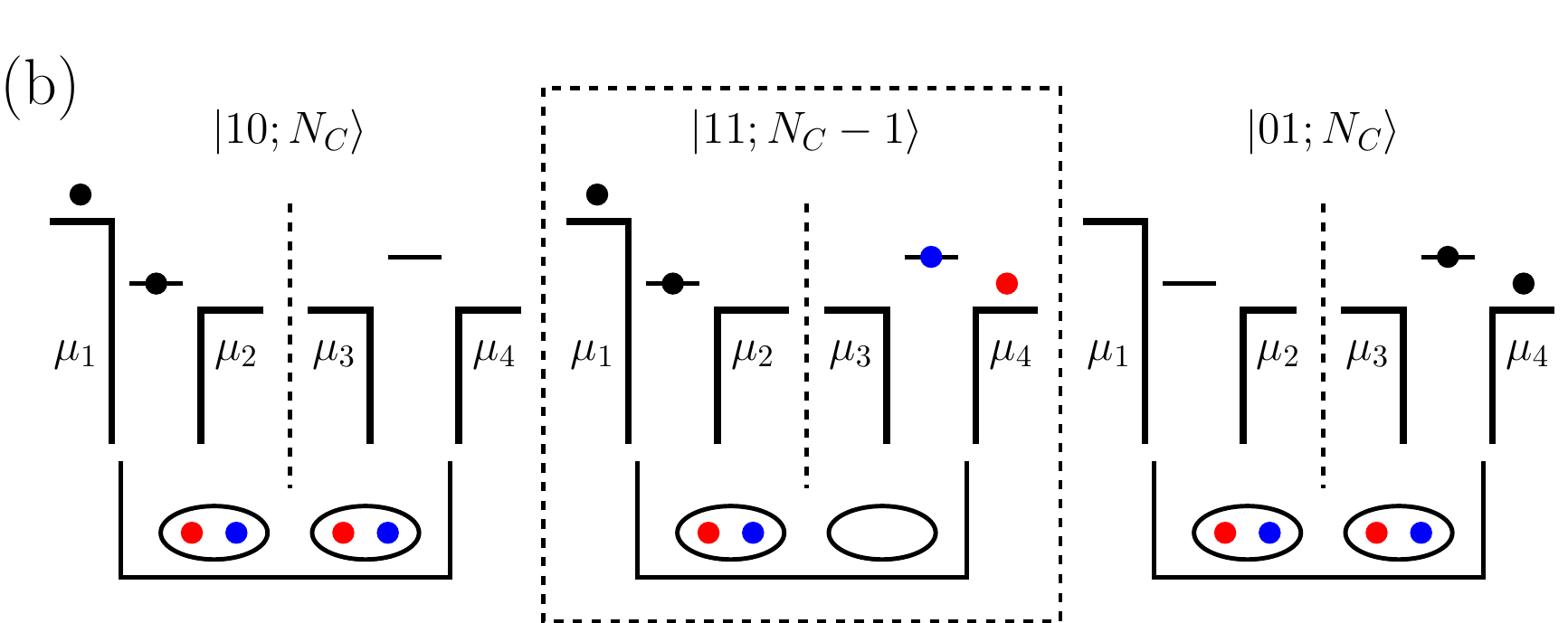}
\caption{Examples of cotunneling processes through a Majorana box. Dashed boxes mark virtual states. (a) Elastic cotunneling: An electron occupying the MBSs is transferred to lead $2$, whereafter an electron from lead $1$ is transferred into the MBSs. The MBS goes from being occupied to empty and once again occupied, while allowing an electron to be transferred from lead $1$ to lead $2$. The process of the island is $\ket{10;N_{C}}\rightarrow\ket{00;N_{C}}\rightarrow\ket{10;N_{C}}$. (b) Inelastic cotunneling: a Cooper pair is broken up whereupon one of the electrons forming the Cooper pair is transferred into lead $3$ or $4$, whereas the other one is left to occupy the MBSs in the second wire. The electron in lead $1$ and the one that occupies the MBSs of wire one are deposited into the superconductor where they form a Cooper pair. The state of the island changes as follows: $\ket{10;N_{C}}\rightarrow\ket{11;N_{C}-1}\rightarrow\ket{01;N_{C}}$. An electron has been transferred from lead $1$ to lead $3$ or $4$.}
\label{Fig: Transport_mech}
\end{figure}

Figure~\ref{Fig: Transport_mech} explains the physical mechanisms behind these transport processes. To be specific, we start in the initial state $\ket{10;0}$. Next, the electron occupying the $d_{1}$ mode is transferred to lead 2, leaving the dot in the virtual state $\ket{00;N_{C}}$. An electron is then transferred from lead 1 onto the island and the system returns to the $\ket{10;N_{C}}$ state. The process is depicted in Fig.~\ref{Fig: Transport_mech}(a), and allows an electron to be transferred from lead 1 to lead 2.

As in the sequential tunneling regime the Cooper pairs enables tunneling between lead 1 and leads 3 and 4. Consider Fig.~\ref{Fig: Transport_mech}(b). If $\mu_{3,4}< -(\xi_{2} - \xi_{1})$ a Cooper pair can be split and one of the electrons will be transferred to lead 3 or 4 and the other one to the $d_{2}$ mode. Thereafter an electron from leads 1 or 2 can, with the electron occupying the $d_{1}$ mode, form a Cooper pair. Thus an electron has been transferred from either lead 1 or 2 to either lead 3 or 4.

\subsubsection{T-junction}
\begin{figure} [t]
\includegraphics[trim=5cm 7.5cm 6cm 6cm,scale=0.4]{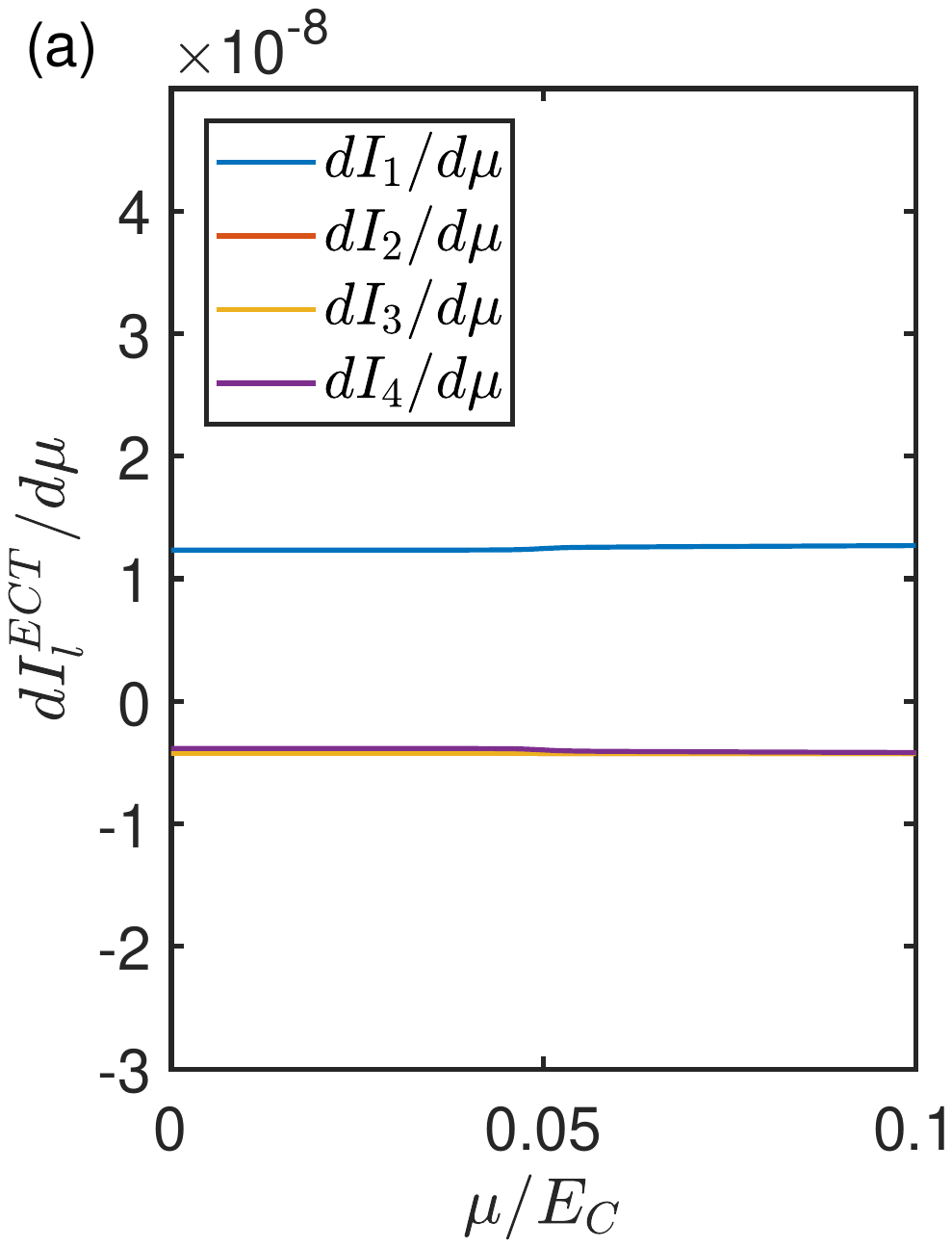}
\includegraphics[trim=5cm 7.5cm 6cm 6cm,scale=0.4]{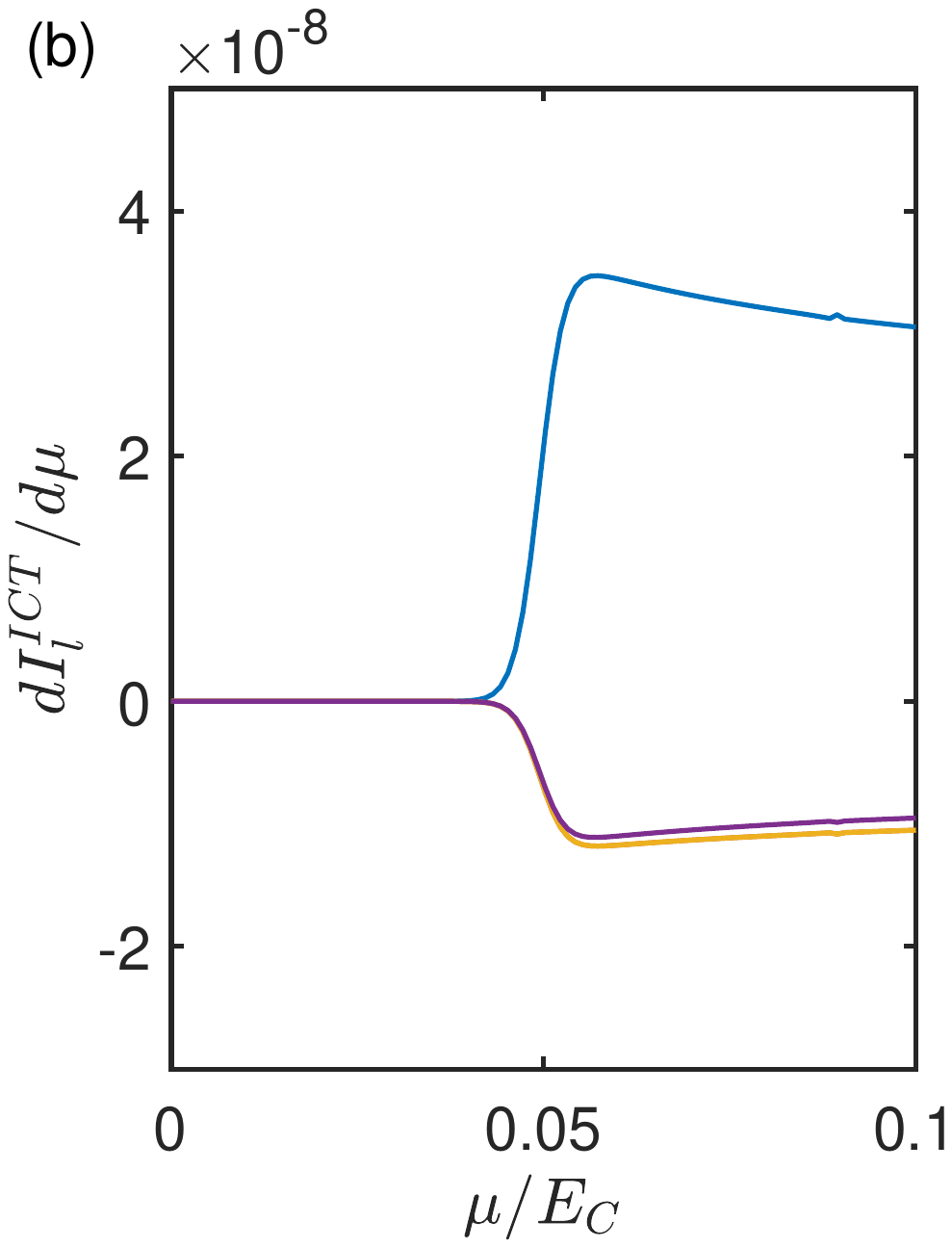}
\caption{ Differential conductance in the cotunneling limit for the T-junction and a symmetric bias configuration, $\mu_{1} = \mu$ and $\mu_{2,3,4} = - \mu$. (a) Differential conductance due to elastic cotunneling. (b) Differential conductance due to inelastic cotunneling. In both plots $dI_{2}/d\mu = dI_{3}/d\mu$. The parameters are $E_{C} = 1$, $W_{i} = 10^{-4}$, $\xi_{T} = 0.05$, $\beta = 900$ and $n_{g} = - 1$.}
\label{Fig: CTT}
\end{figure}

The differential conductance of the T-junction due to cotunneling processes is shown in Fig.~\ref{Fig: CTT}. In most respects the results are very similar to what is observed for the Majorana box. However, in contrast to the Majorana box, one finds a current due to both elastic and inelastic cotunneling in all leads. When $|\mu| < \xi_{T}$, only elastic cotunneling contributes to the current, whereas for $|\mu| > \xi_{T}$, currents due to inelastic cotunneling are also observed.

Compared to the sequential tunneling limit a current between the outer and central leads is possible even for bias voltages smaller than $\xi_{T}$. This is a consequence of the fact that tunneling via virtual states is possible in the cotunneling limit. Suppose the island is in the $\ket{10;0}$ state. An electron tunnels onto the island which is now in the $\ket{11;0}$ state. Since both the central lead and the outer leads couple to the $d_{2}$ mode, an electron can now leave the island in any of the leads. The island then returns to the $\ket{10;0}$ state. This is essentially also the reason for why one finds a current in all the leads for the T-junction but not in the Majorana box: if all leads are connected to the same Dirac state $d_2$, a current will flow between all the leads, regardless of the magnitude of the applied bias.

\section{Conclusions} \label{Sec: Conclusions}

In summary we have studied the transport properties of networks of Majorana bound states, in particular the T-junction and the Majorana box geometry. Assuming that the coupling to the leads is weak, we have investigated the sequential tunneling and the cotunneling limit, and have taken into account both the charging energy and the overlap between the Majorana bound states. We have found that the combination of the latter with the former gives rise to novel transport features.

In the sequential tunneling regime we found currents due to resonant tunneling for both the T-junction and the Majorana box. Using a bias configuration where one lead has a positive chemical potential whereas the three other leads have equal negative biases, we found that for small bias voltages, transport is blocked in the central lead of the T-junction or, respectively, the unbiased wire of the Majorana box. Only once a bias exceeding the overlap energy is applied, a current can flow. The observed current in the unbiased wire of the Majorana box is then due to nonlocal charge transport via the Cooper pairs.

In the cotunneling regime transport is due to elastic and inelastic cotunneling. Contrary to the sequential tunneling regime, transport is possible for small bias voltage in the central lead of the T-junction. However, in the case of the Majorana box, current is still only possible if the applied bias exceeds the overlap of the Majoranas on the unbiased wire. As in the sequential tunneling regime, the Cooper pairs of the superconductor allow for a nonlocal transfer of charge between the two wires. The conductance in the cotunneling regime shows bias voltage features only due to inelastic processes directly related to the finite wave function overlaps between MBSs.

Importantly, our results show that the two structures are rather distinct in their transport properties as soon as the overlap between the Majoranas and the Coulomb charging energy are taken into account. We have provided an understanding for these differences and the mechanisms for the different transport processes that were observed. Our results show that the interplay of charging energy and wave function overlap in Majorana bound state systems gives rise to interesting novel transport phenomena.

\begin{acknowledgments}
JE and TLS acknowledge support by the National Research Fund, Luxembourg under grants ATTRACT 7556175 and CORE 11352881. PR acknowledges financial support from the Hannover-Braunschweig science cooperation QUANOMET and DFG-EXC 2123, Quantum Frontiers.
\end{acknowledgments}

\appendix

\section{Bogoliubov transformation} \label{Apn: BTT}
To express the T-junction Hamiltonian in terms of Dirac operators and in diagonal form, we first write the Majorana operators in terms of Dirac operators after which we perform a Bogoliubov transform. We define [see Eq.~(\ref{Eq: Maj12})]
\begin{align}
\gamma_{1} &= c_1 + c_{1}^{\dagger} \quad \gamma_{2} = i(c_{1}^{\dagger} - c_{1}) \\
\gamma_{3} &= c_2 + c_{2}^{\dagger} \quad \gamma_{4} = i(c_{2}^{\dagger} - c_{2})
\end{align}
By inserting these expressions into Eq.~(\ref{eq:T}), the Hamiltonian can be written in Nambu form, $H_{\rm T-junction} = \frac{1}{2}\mathbf{C}^\dag\mathcal{H}\mathbf{C}$ where $\mathbf{C}^\dag = (c_1^{\dagger}, c_2^{\dagger},c_1,c_2)$ and
\begin{equation}
\mathcal{H} = \epsilon\begin{pmatrix}
0 & -(1+i) & 0& (1+i) \\
-(1-i) & -2 & -(1+i) & 0 \\
0 & -(1-i) & 0 & (1-i) \\
(1-i) & 0 & (1+i) & 2
\end{pmatrix}.
\label{Eq: Nambu}
\end{equation}
We now search for a transformation $T_n$,
\begin{equation}
\mathbf{C} = T_{n}\mathbf{D},
\end{equation}
which diagonalizes the Hamiltonian. The eigenmodes of the system are combined to $\mathbf{D}^\dag = (d_{2}^{\dagger}, d_{1}^{\dagger}, d_{2}, d_{1})$. The first step is to find the eigenvalues of $\mathcal{H}$ and their respective eigenvectors. For every positive eigenvalue $\omega$ of Eq.~(\ref{Eq: Nambu}) with eigenvector $v_i(\omega)$, one can find a corresponding negative eigenvalue for which the eigenvector satisfies \cite{xiao_2009}
\begin{align}
v_{i}(-\omega) &= \Sigma_{x}v_{i}^{*}(\omega), \notag \\
\Sigma_{x} &= \begin{pmatrix}
0&I\\I &0
\end{pmatrix},
\end{align}
where $I$ is the identity matrix. Once the eigenvectors are found $T_{n}$ can be constructed
\begin{equation}
T_{n} = [v(\omega_{1}),\, v(\omega_{2}), \, v(-\omega_{1}), \, v(-\omega_{2})].
\end{equation}
The Hamiltonian matrix $\mathcal{H}$ in Eq.~(\ref{Eq: Nambu}) has two nondegenerate nonzero eigenvalues $\pm 2\sqrt{3}\epsilon = \pm \xi_T$ and a twofold degenerate zero eigenvalue. The corresponding eigenvectors gives us the basis transformation matrix
\begin{equation}
T_{n} = \begin{pmatrix}
x_{1} & x_{0} & ix_{1}^{*} & (-ix_{0}^{*} - \alpha^{*}y_{0}^{*}) \\ \frac{1}{2}\alpha\beta x_{1} & y_{0} & \frac{1}{2}\alpha^{*}\tilde{\beta}x_{1}^{*} & y_{0}^{*} \\ -ix_{1} & (ix_{0} - \alpha y_{0}) & x_{1}^{*} & x_{0}^{*} \\ \frac{1}{2}\alpha \tilde{\beta}x_{1} & y_{0} & \frac{1}{2}\alpha^{*}\beta x_{1}^{*} & y_{0}^{*}
\end{pmatrix}.
\end{equation}
Orthonormality of the eigenvectors corresponds to the following conditions
\begin{align}
&2|x_{0}|^{2} + 4|y_{0}|^{2} + i(1-i)x_{0}^{*}y_{0} - i(1+i)y_{0}^{*}x_{0} = 1, \notag \\
&y_{0}^{2} + ix_{0}^{2} = (1-i)x_{0}y_{0}, \notag \\
&|x_{1}| = \frac{1}{\sqrt{6}}.
\end{align}
By solving these equations and transforming from the diagonal $d_{1}$ and $d_{2}$ basis back to the Majorana basis we obtain the following representation of the MBSs in terms of the eigenstates of $H_{\rm T-junction}$,
\begin{align}
\gamma_{1} &= \frac{1}{\sqrt{3}}d_{1} + \sqrt{\frac{2}{3}}e^{i\pi/12}d_{2} + \hc , \\
\gamma_{2} &= \frac{1}{\sqrt{3}}d_{1} + \sqrt{\frac{2}{3}}e^{-7i\pi/12}d_{2} + \hc, \\
\gamma_{3} &= \frac{1}{\sqrt{3}}d_{1} + \sqrt{\frac{2}{3}}e^{3i\pi/4}d_{2} + \hc, \\
\gamma_{4} &= i(d_{1} - d_{1}^{\dagger}).
\end{align}
This leads to the coefficients $\alpha_{lj}$ listed in Table~(\ref{Tab: Tjunction}).

\section{Second order transition rates} \label{Apn: rates}
We apply Fermi's golden rule to calculate the cotunneling transition rates. Fermi's golden rule reads
\begin{equation}
W^{\alpha}_{\beta} = 2\pi\sum_{i,f}\left| \bra{\psi_{f},\beta} \hat{T} \ket{\psi_{i},\alpha} \right|^{2}\delta(E_{f,\beta} - E_{i,\alpha}),
\end{equation}
with
\begin{equation}
\hat{T} = H_{tun}^{\prime} + H_{tun}^{\prime}\frac{1}{H_{0} - E_{i,\alpha}}\hat{T}.
\end{equation}
The second order term becomes
\begin{align}
W^{\alpha}_{\beta}
 &= 2\pi\sum_{i,f}\left| \bra{\psi_f,\beta} H_{tun}^{\prime}\frac{1}{H_{0} - E_{i,\alpha}}H_{tun}^{\prime} \ket{ \psi_{i},\alpha} \right|^{2} \notag \\
&\times
    \delta(E_{f,\beta} - E_{i,\alpha}).
\end{align}
Here, $E_{i,\alpha}$ and $E_{f,\beta}$ are the initial and final energy, respectively, for a transition from initial state $\ket{\psi_i, \alpha}$ to final state $\ket{\psi_f,\beta}$. The unperturbed Hamiltonian is given by
\begin{equation}
H_{0} = H_{\rm leads} + H_{\rm MBS} + H_{\rm charging}.
\end{equation}
Taking the leads to follow a Fermi-distribution and writing out the tunneling Hamiltonian we obtain all the different processes allowed in second order perturbation theory. We have furthermore assumed a constant density of states, $D_{l} = 1/(2 \pi v_F)$, in the leads and defined $W_{l} = 2\pi|t_{l}|^{2}D_{l}$. We obtain
\begin{widetext}
\small
\begin{align*}
W^{11;-1}_{11;0}(l_{1},l_{2}) &= \frac{W_{l_{1}}W_{l_{2}}}{2\pi}\int dE n_{F}(E - \mu_{l_{1}})n_{F}(-E + E_{C}(4+4n_{g}) - \mu_{l_{2}}) \\
& \times \Big| \frac{\alpha_{l_{1}1}^{*}\alpha_{l_{2}1}}{\xi_{1} + E_{C}(3+2n_{g}) -E}  - \frac{\alpha_{l_{2}1}^{*}\alpha_{l_{1}1}}{\xi_{1} - E_{C}(1+2n_{g}) + E} + \frac{\alpha_{l_{1}2}^{*}\alpha_{l_{2}2}}{\xi_{2} + E_{C}(3+2n_{g}) - E} - \frac{\alpha_{l_{2}2}^{*}\alpha_{l_{1}2}}{\xi_{2} - E_{C}(1 + 2n_{g}) + E} \Big|^{2}
\end{align*}

\begin{align*}
W^{11;0}_{11;-1}(l_{1},l_{2}) &= \frac{W_{l_{1}}W_{l_{2}}}{2\pi}\int dE \left[ 1 - n_{F}(E - \mu_{l_{1}})\right]\left[ 1 - n_{F}(-E + E_{C}(4 + 4n_{g}) - \mu_{l_{2}})\right] \\
& \times \Big| \frac{\alpha_{l_{1}1}^{*}\alpha_{l_{2}1}}{\xi_{1} - E_{C}(1 + 2n_{g}) + E }   - \frac{\alpha_{l_{2}1}^{*}\alpha_{l_{1}1}}{\xi_{1} + E_{C}(3 + 2n_{g}) - E} + \frac{\alpha_{l_{1}2}^{*}\alpha_{l_{2}2}}{\xi_{2} - E_{C}(1 + 2n_{g}) + E} - \frac{\alpha_{l_{2}2}^{*}\alpha_{l_{1}2}}{\xi_{2} + E_{C}(3 + 2n_{g}) - E} \Big|^{2}
\end{align*}

\begin{align*}
W^{00;0}_{00;1}(l_{1},l_{2}) &= \frac{W_{l_{1}}W_{l_{2}}}{2\pi}\int dEn_{F}(E - \mu_{l_{1}})n_{F}( -E + E_{C}(4+4n_{g}) - \mu_{l_{2}}) \\
& \times \Big| \frac{\alpha_{l_{1}1}\alpha_{l_{2}1}^{*}}{-\xi_{1} + E_{C}(3+2n_{g}) -E}  - \frac{\alpha_{l_{2}1}\alpha_{l_{1}1}^{*}}{-\xi_{1} - E_{C}(1+2n_{g}) + E} + \frac{\alpha_{l_{1}2}\alpha_{l_{2}2}^{*}}{-\xi_{2} + E_{C}(3 + 2n_{g}) - E} - \frac{\alpha_{l_{2}2}\alpha_{l_{1}2}^{*}}{-\xi_{2} - E_{C}(1 + 2n_{g}) + E} \Big|^{2}
\end{align*}

\begin{align*}
W^{00;1}_{00;0}(l_{1},l_{2}) &= \frac{W_{l_{1}}W_{l_{2}}}{2\pi}\int dE \left[ 1 - n_{F}(E - \mu_{l_{1}})\right]\left[ 1 - n_{F}(-E + E_{C}(4+4n_{g}) - \mu_{l_{2}})\right] \\
& \times \Big| \frac{\alpha_{l_{1}1}\alpha_{l_{2}1}^{*}}{-\xi_{1} - E_{C}(1 + 2n_{g}) + E}  - \frac{\alpha_{l_{2}1}\alpha_{l_{1}1}^{*}}{-\xi_{1} + E_{C}(3+2n_{g}) - E} + \frac{\alpha_{l_{1}2}\alpha_{l_{2}2}^{*}}{-\xi_{2} -E_{C}(1 + 2n_{g}) + E} - \frac{\alpha_{l_{2}2}\alpha_{l_{1}2}^{*}}{-\xi_{2} + E_{C}(3+2n_{g}) - E} \Big|^{2}
\end{align*}

\begin{align*}
W^{01;0}_{10;0}(l_{1},l_{2}) &= \frac{W_{l_{1}}W_{l_{2}}}{2\pi}\int dE n_{F}(E - \mu_{l_{1}})\left[ 1 - n_{F}(E + \xi_{2} - \xi_{1} - \mu_{l_{2}})\right] \\
& \times \Big| \frac{\alpha_{l_{2}2}\alpha_{l_{1}1}^{*}}{-\xi_{1} - E_{C}(3+2n_{g}) + E}  - \frac{\alpha_{l_{1}2}\alpha_{l_{2}1}^{*}}{-\xi_{2} + E_{C}(1+2n_{g}) - E} + \frac{\alpha_{l_{1}1}^{*}\alpha_{l_{2}2}}{2\xi_{2} - \xi_{1} - E_{C}(3 + 2n_{g}) + E} - \frac{\alpha_{l_{2}1}^{*}\alpha_{l_{1}2}}{\xi_{2} + E_{C}(1 + 2n_{g}) - E} \Big|^{2}
\end{align*}

\begin{align*}
W^{10;0}_{01;0}(l_{1},l_{2}) &= \frac{W_{l_{1}}W_{l_{2}}}{2\pi}\int dE n_{F}(E - \mu_{l_{1}})\left[ 1 - n_{F}(E + \xi_{1} -\xi_{2} - \mu_{l_{2}})\right] \\
& \times \Big| \frac{\alpha_{l_{2}1}\alpha_{l_{1}2}^{*}}{-\xi_{2} - E_{C}(3+2n_{g}) + E}  - \frac{\alpha_{l_{1}1}\alpha_{l_{2}2}^{*}}{-\xi_{1} + E_{C}(1+2n_{g}) - E} + \frac{\alpha_{l_{1}2}^{*}\alpha_{l_{2}1}}{2\xi_{1}-\xi_{2} - E_{C}(3+2n_{g}) + E} - \frac{\alpha_{l_{2}2}^{*}\alpha_{l_{1}1}}{\xi_{1} + E_{C}(1+2n_{g}) - E} \Big|^{2}
\end{align*}

\begin{align*}
W^{11;0}_{00;1}(l_{1},l_{2}) &= \frac{W_{l_{1}}W_{l_{2}}}{2\pi}\int dE n_{F}(E - \mu_{l_{1}})\left[ 1 - n_{F}(E + \xi_{1} + \xi_{2} - \mu_{l_{2}})\right] \\
& \times \Big| \frac{\alpha_{l_{2}1}\alpha_{l_{1}2}}{\xi_{2} - E_{C}(5 + 2n_{g}) + E}  - \frac{\alpha_{l_{1}1}\alpha_{l_{2}2}}{-\xi_{1} + E_{C}(3 + 2n_{g}) - E} + \frac{\alpha_{l_{2}2}\alpha_{l_{1}1}}{\xi_{1} - E_{C}(5 + 2n_{g})  + E} - \frac{\alpha_{l_{1}2}\alpha_{l_{2}1}}{-\xi_{2} + E_{C}(3+2n_{g}) - E} \Big|^{2}
\end{align*}

\begin{align*}
W^{00;1}_{11;0}(l_{1},l_{2}) &= \frac{W_{l_{1}}W_{l_{2}}}{2\pi}\int dE n_{F}(E - \mu_{l_{1}})\left[ 1 - n_{F}(E -\xi_{1} - \xi_{2} - \mu_{l_{2}})\right] \\
& \times \Big| \frac{\alpha_{l_{1}1}^{*}\alpha_{l_{2}2}^{*}}{\xi_{1} + E_{C}(3 + 2n_{g}) - E}  - \frac{\alpha_{l_{2}1}^{*}\alpha_{l_{1}2}^{*}}{-\xi_{2} - E_{C}(5 + 2n_{g}) + E} + \frac{\alpha_{l_{1}2}^{*}\alpha_{l_{2}1}^{*}}{\xi_{2} + E_{C}(3+2n_{g})  - E} - \frac{\alpha_{l_{2}2}^{*}\alpha_{l_{1}1}^{*}}{-\xi_{1} - E_{C}(5 + 2n_{g}) + E} \Big|^{2}
\end{align*}

\begin{align*}
W^{11;-1}_{00;0}(l_{1},l_{2}) &= \frac{W_{l_{1}}W_{l_{2}}}{2\pi}\int dE n_{F}(E - \mu_{l_{1}})\left[ 1 - n_{F}(E + \xi_{1} + \xi_{2} - \mu_{l_{2}})\right] \\
& \times \Big| \frac{\alpha_{l_{2}1}\alpha_{l_{1}2}}{\xi_{2} - E_{C}(1 + 2n_{g}) + E}  - \frac{\alpha_{l_{1}1}\alpha_{l_{2}2}}{-\xi_{1} - E_{C}(1 - 2n_{g}) - E} + \frac{\alpha_{l_{2}2}\alpha_{l_{1}1}}{\xi_{1} - E_{C}(1 + 2n_{g})  + E} - \frac{\alpha_{l_{1}2}\alpha_{l_{2}1}}{-\xi_{2} - E_{C}(1 - 2n_{g}) - E} \Big|^{2}
\end{align*}

\begin{align*}
W^{00;0}_{11;-1}(l_{1},l_{2}) &= \frac{W_{l_{1}}W_{l_{2}}}{2\pi}\int dE n_{F}(E - \mu_{l_{1}})\left[ 1 - n_{F}(E  -\xi_{1} - \xi_{2} - \mu_{l_{2}})\right] \\
& \times \Big| \frac{\alpha_{l_{1}1}^{*}\alpha_{l_{2}2}^{*}}{\xi_{1} - E_{C}(1 - 2n_{g}) - E}  - \frac{\alpha_{l_{2}1}^{*}\alpha_{l_{1}2}^{*}}{-\xi_{2} - E_{C}(1 + 2n_{g}) + E} + \frac{\alpha_{l_{1}2}^{*}\alpha_{l_{2}1}^{*}}{\xi_{2} - E_{C}(1-2n_{g})  - E} - \frac{\alpha_{l_{2}2}^{*}\alpha_{l_{1}1}^{*}}{-\xi_{1} - E_{C}(1 + 2n_{g}) + E} \Big|^{2}
\end{align*}

\begin{align*}
W^{11;-1}_{00;1}(l_{1},l_{2}) &= \frac{W_{l_{1}}W_{l_{2}}}{2\pi}\int dE n_{F}(E - \mu_{l_{1}})n_{F}(-E -\xi_{1} - \xi_{2} + E_{C}(4 +4n_{g}) - \mu_{l_{2}}) \\
& \times \Big| \frac{\alpha_{l_{2}1}\alpha_{l_{1}2}}{\xi_{2} - E_{C}(1 + 2n_{g}) + E}  - \frac{\alpha_{l_{1}1}\alpha_{l_{2}2}}{-\xi_{1} + E_{C}(3+2n_{g}) - E} + \frac{\alpha_{l_{2}2}\alpha_{l_{1}1}}{\xi_{1} - E_{C}(1 + 2n_{g}) + E} - \frac{\alpha_{l_{1}2}\alpha_{l_{2}1}}{-\xi_{2} + E_{C}(3+2n_{g}) - E} \Big|^{2}
\end{align*}

\begin{align*}
W^{00;1}_{11;-1}(l_{1},l_{2}) &= \frac{W_{l_{1}}W_{l_{2}}}{2\pi}\int dE \left[1-n_{F}(E - \mu_{l_{1}})\right]\left[1 - n_{F}(-E -\xi_{1} - \xi_{2} + E_{C}(4 + 4n_{g}) - \mu_{l_{2}})\right] \\
& \times \Big| \frac{\alpha_{l_{2}1}^{*}\alpha_{l_{1}2}^{*}}{-\xi_{2} + E_{C}(3 + 2n_{g}) - E}  - \frac{\alpha_{l_{1}1}^{*}\alpha_{l_{2}2}^{*}}{\xi_{1} - E_{C}(1 + 2n_{g}) + E} + \frac{\alpha_{l_{2}2}^{*}\alpha_{l_{1}1}^{*}}{-\xi_{1} + E_{C}(3 + 2n_{g}) - E} - \frac{\alpha_{l_{1}2}^{*}\alpha_{l_{2}1}^{*}}{\xi_{2} - E_{C}(1 + 2n_{g}) + E} \Big|^{2}
\end{align*}

\begin{align*}
W^{00;0}_{11;0}(l_{1},l_{2}) &= \frac{W_{l_{1}}W_{l_{2}}}{2\pi}\int dE n_{F}(E - \mu_{l_{1}})n_{F}(-E + \xi_{1} + \xi_{2} + E_{C}(4 + 4n_{g}) - \mu_{l_{2}}) \\
& \times \Big| \frac{\alpha_{l_{2}1}^{*}\alpha_{l_{1}2}^{*}}{-\xi_{2} - E_{C}(1+2n_{g}) + E}  - \frac{\alpha_{l_{1}1}^{*}\alpha_{l_{2}2}^{*}}{\xi_{1} + E_{C}(3 + 2n_{g}) - E} + \frac{\alpha_{l_{2}2}^{*}\alpha_{l_{1}1}^{*}}{-\xi_{1} - E_{C}(1 + 2n_{g}) + E} - \frac{\alpha_{l_{1}2}^{*}\alpha_{l_{2}1}^{*}}{\xi_{2} + E_{C}(3 + 2n_{g}) - E} \Big|^{2}
\end{align*}

\begin{align*}
W^{11;0}_{00;0}(l_{1},l_{2}) &= \frac{W_{l_{1}}W_{l_{2}}}{2\pi}\int dE \left[1-n_{F}(E - \mu_{l_{1}})\right]\left[1 - n_{F}(-E + \xi_{1} + \xi_{2} + E_{C}(4 + 4n_{g}) - \mu_{l_{2}})\right] \\
& \times \Big| \frac{\alpha_{l_{2}1}\alpha_{l_{1}2}}{\xi_{2} + E_{C}(3 + 2n_{g}) - E}  - \frac{\alpha_{l_{1}1}\alpha_{l_{2}2}}{-\xi_{1}-E_{C}(1 + 2n_{g}) + E} + \frac{\alpha_{l_{2}2}\alpha_{l_{1}1}}{\xi_{1} + E_{C}(3 + 2n_{g}) - E} - \frac{\alpha_{l_{1}2}\alpha_{l_{2}1}}{-\xi_{2} - E_{C}(1+2n_{g}) + E} \Big|^{2}.
\end{align*}
\end{widetext}
\normalsize

\bibliography{bibliography}

\end{document}